# The Quality of the 2020 Census

*An Independent Assessment of Census Bureau Activities Critical to Data Quality*

PAUL BIEMER
JOSEPH SALVO
JONATHAN AUERBACH

# Contents

# I Executive Summary

## Purpose of This Analysis

The challenges facing the 2020 census were unprecedented. The coronavirus outbreak hit just as the Census Bureau began its mail-out procedures. The census schedule, encompassing hundreds of interconnected activities, changed repeatedly as conditions worsened. Activities were modified or canceled in real time to accommodate schedule changes, only to have the schedule revised again. Wildfires in the West and weather events in the South further impeded operations. Because of these and other challenges, many data users voiced concerns about potential coverage errors. Such errors jeopardize the fair division of representation and resources, the basis of U.S. government outlined in the Constitution.

These concerns prompted the American Statistical Association (ASA) to convene the 2020 Census Quality Indicators Task Force, whose deliberations culminated in an October 2020 report, *Census Quality Indicators: A Report from the American Statistical Association*. The report recommended that the Census Bureau grant external census experts access to operational data so they can independently address the concerns of data users. The Census Bureau agreed, and the Task Force selected Paul Biemer, Joseph Salvo, and Jonathan Auerbach to conduct an evaluation.

This document summarizes the findings of this independent evaluation of the quality of the state-level population counts released for congressional apportionment. It examines 10 process statistics (PSs) largely based on the quality indicators recommended in the Task Force report. Each PS reflects an activity or group of activities in the Bureau's schedule that may have been affected by the pandemic, other unprecedented events, or otherwise present some appreciable risk of producing errors in the 2020 census count as determined by the authors.

This document does not consider the quality of characteristics, such as sex, race, or ethnicity, or the quality of the population count at smaller geographies, such as blocks, tracts, or counties. Further, because of the limitations of the data currently available, it does not attempt to quantify the impact of coverage errors on congressional apportionment.

These findings contribute to the ongoing discussion about the quality of the 2020 census. Data users may also wish to review the growing body of literature on 2020 census quality to supplement the information in this report. For example, the Census Bureau engaged the JASON and the National Academies' Committee on National Statistics to assess 2020 census quality. The Bureau also released its own preliminary evaluation in April 2020, and it will soon release the Post Enumeration Survey, Demographic Analysis, and the 2020 Census Program for Evaluations, Assessments, and Experiments.

## Summary of Methodology

This analysis partitions the census enumeration process into five operational phases and identifies major activities within each phase that are critical to 2020 census data quality. For each critical activity, one or more PSs are defined at the state level to assess the performance of the activity as it may affect the apportionment counts. All PSs defined in this report are proportions of cases (i.e., housing units [HUs], persons or addresses) affected by the operation within its particular universe or purview. For example, one critical activity considered is the imputation of HU status. The PS defined for this activity is the proportion of addresses in a state for which HU status was imputed.

Many activities associated with the census are interrelated, including the critical activities identified in this analysis. For example, administrative records are sometimes used in place of proxy respondents. In addition, field verification and quality assurance checks are conducted continually throughout the enumeration and are designed to reduce the risk of coverage error. No assessment of the effectiveness of these checks is available and will therefore not be considered in this analysis.

A total of 10 critical activities and associated PSs are defined for the 2020 census. Six of these compare the same critical activity from the 2020 and 2010 censuses. Thus, these PSs reflect the change in the frequency a critical activity was employed in 2020 relative to 2010. In four cases, the critical activity either did not exist in 2010 or its comparability to 2020 could not be established. Thus, these PSs only reflect the critical activity employed in 2020.

Assuming the PS is an indicator of coverage error risk from the critical activity, the variation in the PSs across states reflects the variation in error risks from that activity. Thus, for PSs that are differences between 2020 and 2010, positive values imply greater risk from the activity in 2020 than in 2010 while negative values imply lesser risk in 2020. Similarly, for PSs defined for 2020 only, larger values for a state imply larger risks from the critical activity compared to states with smaller values.

## Summary of Main Findings

The analysis in this report finds substantial increases in the frequency of several critical activities believed to be associated with higher error risk by the authors. Main findings can be divided into three groups: (1) Increase in critical activities from 2010 to 2020 that were expected, and Census Bureau procedures mitigated some of the error risk; (2) Increase in critical activities from 2010 to 2020 that were not fully expected, and additional procedures may not have mitigated error risk; and (3) Critical activities whose use in 2020 varied considerably across states.

1. *Increase in critical activities from 2010 to 2020 that were expected, and Census Bureau procedures mitigated some of the error risk.*

Households for which the Census Bureau obtained multiple responses—either by enumeration or other census activity—increased by nearly 20 percentage points in 2020 compared to 2010. Part of the increase in multiple responses is the result of allowing households to respond over the Internet, which was a new response option in 2020. The ease of responding by Internet likely increased the number of responses per household. Anticipating this increase and its associated error risks, the Bureau established a number of auxiliary error mitigation procedures. For example, the Bureau prepared a quality control approach referred to as the Primary Selection Algorithm to select the best response from the multiple responses, which at least partially mitigated the risk of overcounting.

The use of administrative records was a new critical activity in 2020. Administrative records were used to enumerate up to 5% of households in some states. An error risk associated with its use is that records may be outdated or otherwise inaccurate, which may translate into coverage errors. This risk was mitigated to some extent through the use of multiple sources of administrative data rather than a single source.

*2. Increase in critical activities from 2010 to 2020 that were not fully expected, and additional procedures may not have mitigated error risk.*

The frequency that college students were initially enumerated in wrong locations also increased in 2020 compared to 2010. Although the Bureau has corrective measures in place to mitigate these errors, it could not have anticipated the pandemic and its effect on school closings, which was to displace greater numbers of college students away from their usual residences. Although the Bureau had established procedures for dealing with persons not counted at the usual place of residence, the increased number of displacements increased the potential for error from those procedures.

*3. Critical activities whose use in 2020 varied considerably across states.*

The imputation of persons in Group Quarters (GQs) was a new critical activity in 2020. The percent of all persons in GQs who were imputed varied more than 10 percentage points across states.

Another main finding is that about 11% of addresses were either added to or deleted from the Master Address File (MAF) during the 2020 census. This percentage varied across states by more than 20 points. Because of changes to the MAF development process implemented in 2020 (described subsequently), the volume of MAF revisions likely exceeds those made to the MAF in 2010; however, there are no data available to verify this.

**The Larger Picture**

As recommended by the ASA Task Force report, the findings in this report are all based on PSs calculated from operational data and are subject to a number of limitations. First, the PS reflects the frequency of an activity that could result in an error and not actual errors. As an example, the Bureau attempts to identify and reassign all college students initially counted in the wrong place to their proper locations. Although there is an error risk associated with this activity, the extent of the error is unknown. Without that information, it is impossible to draw conclusions about the error contributed by the activity. Second, errors introduced by one activity could be either mitigated or exacerbated by other activities. In that regard, the PSs may overstate or understate the actual error risk for an activity. Third, the selection of critical activities for analysis is a subjective process and based on the knowledge and experience of the authors. It is conceivable that another team of researchers could select other PSs that would lead to different determinations about 2020 census quality.

Nevertheless, the analysis presented in this report is an essential first step towards better understanding 2020 Census data quality. It provides important context and insights that will facilitate the interpretation of the 2020 Census Post-Enumeration Survey estimates of undercounts and overcounts that will be released in early 2022.

The strongest statement that can be made based on this analysis is that *most of the critical activities considered in this report were exercised more in 2020 than in 2010.* To the extent these activities reflect coverage error risk, it follows that the risk of coverage errors also increased in 2020 for these activities. However, given the data at hand, this review did not find conclusive evidence that state-level counts used for apportionment purposes are of lower quality in 2020 than in 2010. Nor is there evidence that the apportionment count for any given state is in error.

A more conclusive assessment of 2020 census quality would be achieved by combining the PSs described in this analysis with the results of other methodologies, such as the estimate of undercounts and overcounts from the Census Bureau's Post-Enumeration Survey and the results of the Demographic Analysis.

---

## II Disclaimers

- Process statistics are largely based on the American Statistical Association 2020 Census Quality Indicators Task Force report, modified to reflect Census Bureau limitations and the best judgments of the authors.
- Process statistics reflect the reliance on activities that are believed to increase the risk of an error, not the rate of error. Additional actions taken by the Census Bureau may have mitigated the error risks to an extent that cannot be determined from the data available.
- Both the American Statistical Association 2020 Census Quality Indicators Task Force and the Census Bureau reviewed an earlier version of this report. This version of the report incorporates changes made as a result of those reviews, correcting factual errors and adding additional clarifications. All changes are made at the discretion of the authors.

## III Methodology at a Glance

This analysis partitions the census enumeration process into five phases:

1. **Master Address File (MAF) Development** – update the initial list of residential addresses used by the Census Bureau to contact households and elicit their response to the census.
2. **Self-Response** – encourage a household resident to complete the questionnaire by mail, telephone, or internet.
3. **Nonresponse Follow-up (NRFU)** – a resident of each household that did not self-respond is contacted by a census enumerator who assists in completing the questionnaire.
4. **Data Compilation and Processing** – use statistical methods to fill missing information and resolve multiple responses from the same address.
5. **Group Quarters** – enumerate persons in group facilities such as student housing, nursing homes, and military barracks.

The quality of data from the enumeration process is investigated in three steps. First, an ideal scenario is envisioned for each phase. For example, for MAF Development: Ideally, the MAF would be completely accurate and not require revision. For Self-Response: Ideally, one and only one resident from each household would complete a census questionnaire, etc.

Second, Census Bureau activities that deviate from the ideal scenario are identified. For example, for MAF Development: revising an address on the MAF. For Self-Response: collecting multiple responses for the same household, etc.

Finally, ten process statistics (PSs) are created for all states, DC, and the United States overall, which measure the relative frequency that an activity was exercised, or the opportunity for error from the activity. A higher PS value for a state indicates that a greater use of the activity increased the risk of error in a state's count (see Disclaimers).

As an example, PS 3 in Table 1 measures how often the Census Bureau received more than one response for a household in the 2020 census relative to the 2010 census. Multiple responses have the potential for introducing error because the Bureau must decide which response to count. An error in this decision process could increase coverage error. Thus, greater numbers of HUs with multiple responses imply a greater error risk from this activity.

This analysis converts each PS to a number between 1 and 5 (quintile) where 1 denotes that it ranks in the lower 20% of all states, 2 denotes that it ranks between 20% and 40% of all states, and so on. The higher a state's rank, the higher the risk from that activity. A summary statistic is created by taking the weighted average of the quintile ranks for the 10 PSs, where the weights are proportional to the number of cases within the universe or purview of the activity. This summary process statistic (SPS) ranges from 1 to 5 and is regarded as an average error risk from all 10 critical activities.

## IV Process Statistics Profiles

This section provides an overview of each Process Statistic (PS). It begins by reviewing the five phases of census operations. It then explains how the statistics relate to these phases. This is followed by a profile of each statistic including a combined statistic referred to as the summary process statistic (SPS). The first page of each profile reviews how the statistic is calculated, how it is interpreted, key points, and future research. A choropleth map displays the potential risk implied by the statistic for each state. The second page contains a table with the value of the statistic for each state, including Washington D.C., and the United States overall. The states are sorted by the potential risk reflected by the statistic—from the lowest risk state (rank 1) to the highest (rank 51). Six PSs are the difference in percentages between the 2020 and 2010 censuses. For these, the table contains an additional column showing the 2020 census percentage for reference purposes. A bar chart displays how the statistic varies by state. For the lowest ranked state, the bar is of 0 length and for the highest ranked state, the bar fills the entire cell. Both the map and the table are shaded according to the amount of error risk implied by the PS. Darker shades denote higher risks and lighter shades denote lower risks.

The profiles use Census Bureau terminology. The most common terms are defined below, using abridged definitions from the Census Bureau's Glossary. See the full glossary (census.gov/glossary/) for unabridged definitions or terms not defined here.

**Administrative Record (Admin Rec)** – Records collected by Federal, Tribal, State, and local governments while they are administering programs or providing services.

**Group Quarters** – The Census Bureau classifies the large majority of persons not living in housing units as living in GQs. GQs include such places as college residence halls, residential treatment centers, skilled nursing facilities, group homes, military barracks, correctional facilities, and workers' dormitories.

**Housing Unit** – A house, apartment, trailer, group of rooms, or single room occupied as separate living quarters. If vacant, intended for occupancy as separate living quarters.

**Imputation** – The process used to estimate missing or unacceptable data.

**Master Address File** – The Census Bureau's official inventory of known living quarters (HUs and GQ facilities) and selected nonresidential units (public, private, and commercial).

**Nonresponse Follow-up** – An operation whose objective is to obtain complete survey information from HUs for which the Census Bureau did not receive a completed questionnaire online, by telephone or by mail.

**Self-Response** – The primary method for counting the population. A member of each HU completes the questionnaire online, by telephone or by mail.

## Process Statistics by Census Phase

*Table 1. Process Statistics by Census Phase*

| Process Statistic | | Description |
|---|---|---|
| **MAF Development (MAF)** | | |
| 1 | MAF Revisions | Percent of all addresses that were either deleted or added during the 2020 census data collection period |
| **Self-Response (SR)** | | |
| 2 | Questionnaires Without Identification (ID) not on MAF (Non-Matching No IDs) | Percent of housing units (HUs) submitting questionnaires without census IDs and no matching address was found on the MAF for 2020 |
| 3 | Multiple Responses | Percent of occupied HUs with two or more responses from various sources for 2020 minus the corresponding percentage for 2010 |
| 4 | Usual Residence at College (URC) | Percent of occupied HUs with two or more people where one or more occupant indicated their usual residence was at college for 2020 minus the corresponding percentage for 2010 |
| **Nonresponse Follow-up (NRFU)** | | |
| 5 | Responses Obtained by Proxy (Proxy) | Percent of persons in occupied HUs whose count was obtained by proxy interview for 2020 minus the corresponding percentage for 2010 |
| 6 | Enumerations With Only a Population Count (Count Only) | Percent of occupied HUs where only a population count was obtained for 2020 minus the corresponding percentage for 2010 |
| 7 | Enumerations via Administrative Records (Admin Recs) | Percent of occupied HUs enumerated by administrative records for 2020 |
| **Data Processing** | | |
| 8 | MAF Addresses Having Imputed Status (Status Imputation) | Percent of MAF units whose status was imputed for 2020 minus the corresponding percentage for 2010 |
| 9 | Occupied Housing Units With Imputed Population Counts (Count Imputation) | Percent of occupied HUs with known status but whose population count was imputed for 2020 minus the corresponding percentage for 2010 |
| **Group Quarters (GQs)** | | |
| 10 | Group Quarters With Imputed Count (GQ Imputation) | Percent of the GQs population that was imputed in 2020 |

*Statistics in the text were approved for public dissemination by the Census Bureau Disclosure Review Board. Clearance numbers: CBDRB-FY21-DSSD007-0021, CBDRB-FY21-DSSD007-0024, CBDRB-FY21-DSSD007-0026.

## How the U.S. Population was Counted by Census Phase

Each person below represents one five-hundredth of the occupied housing units counted in the 2020 census (roughly 250,000 households), colored by the phase in which they were counted (see census phases on facing page). MAF not applicable and GQs excluded.

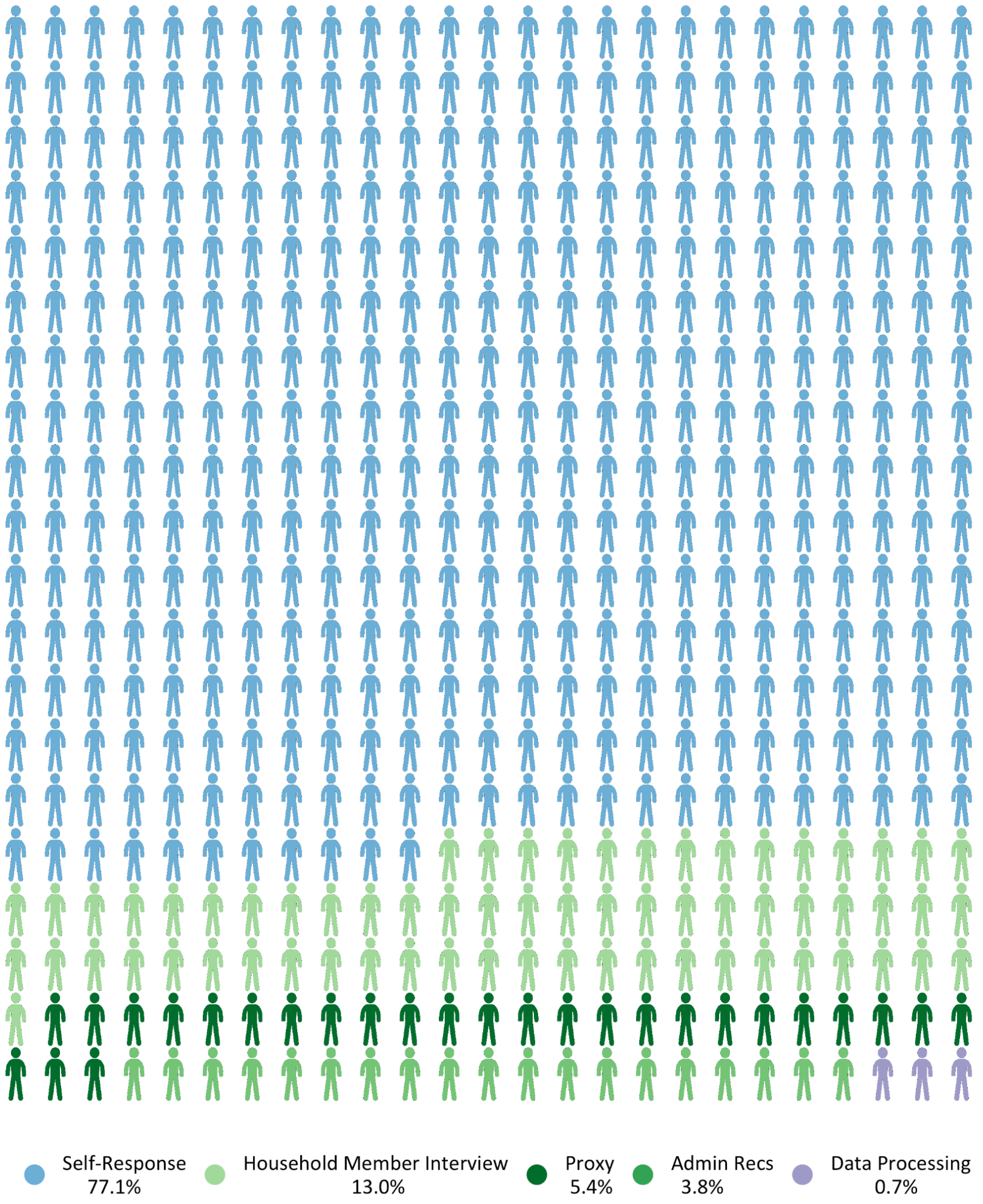

### Process Statistic 1: MAF Revisions

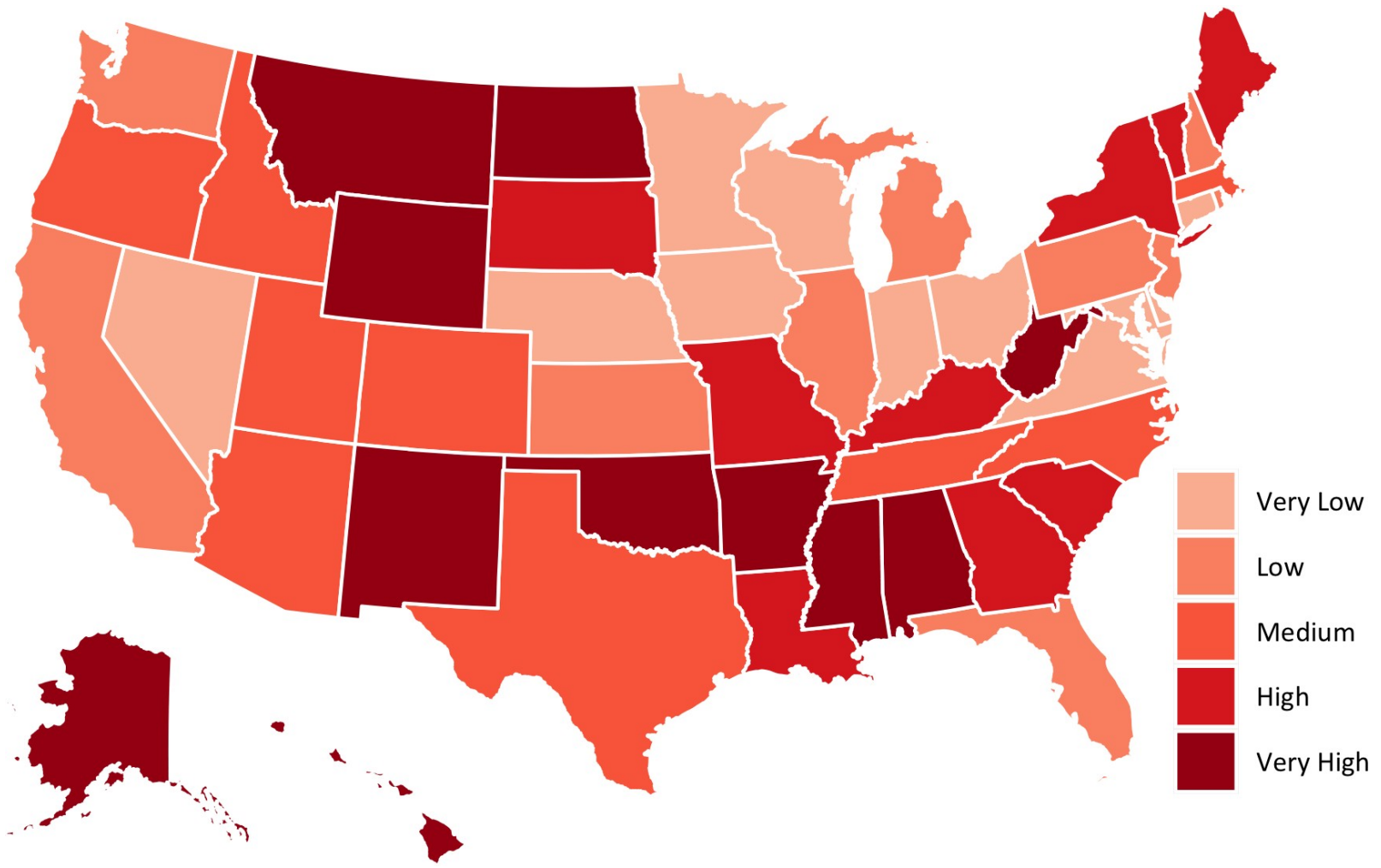

**How is this process statistic calculated?**
Percent of all addresses in the 2020 census that were either deleted or added to the MAF during the data collection period.

**How is this process statistic interpreted?**
Ideally, the address list used by the Census Bureau would not change during the census data collection period. But inevitably some addresses initially on the list are invalid (commercial units, demolished HUs, etc.) and must be deleted by the Bureau, while other addresses initially absent from the list must be added (e.g., new construction, hidden residences). This statistic reflects quality because revision carries risk: invalid addresses may be added or valid addresses deleted. A large volume of revisions suggests that MAF quality is questionable.

**Key Points**

- The percent of addresses that were revised varied widely by state, from 5.9% (MD) to 26% (AK and WV).
- Revisions are highest in largely rural states and "update leave areas" where mail is typically not received at home.
- MAF development activities aimed at verifying many presumably invalid addresses may have increased revisions.

**Future Research**

- Do MAF revisions include invalid addresses/exclude valid ones?
- Did new in-office procedures for resolving addresses increase revisions?
- How might decisions about which addresses to include on the MAF be better informed?

| Rank | State | Process Statistic 1: Percent of all addresses in the census that were either deleted or added to the MAF during the census data collection period for 2020* |
|---|---|---|
| 1 | Maryland | 5.87 |
| 2 | Delaware | 6.24 |
| 3 | Minnesota | 6.48 |
| 4 | Virginia | 6.75 |
| 5 | Wisconsin | 7.07 |
| 6 | Ohio | 7.19 |
| 7 | Indiana | 7.25 |
| 8 | Iowa | 7.27 |
| 9 | Connecticut | 7.56 |
| 10 | Nevada | 7.79 |
| 11 | Nebraska | 7.79 |
| 12 | Washington | 7.80 |
| 13 | Rhode Island | 7.83 |
| 14 | Michigan | 7.89 |
| 15 | Florida | 7.91 |
| 16 | Kansas | 7.98 |
| 17 | New Jersey | 8.33 |
| 18 | Illinois | 8.71 |
| 19 | New Hampshire | 8.79 |
| 20 | California | 8.82 |
| 21 | Pennsylvania | 8.83 |
| | United States | 9.22 |
| 22 | Colorado | 9.95 |
| 23 | Massachusetts | 10.18 |
| 24 | Oregon | 10.20 |
| 25 | Tennessee | 10.29 |
| 26 | North Carolina | 10.55 |
| 27 | Arizona | 10.65 |
| 28 | Utah | 11.01 |
| 29 | Idaho | 11.17 |
| 30 | Texas | 11.25 |
| 31 | Washington, D.C. | 11.51 |
| 32 | Maine | 11.90 |
| 33 | Kentucky | 11.91 |
| 34 | Georgia | 12.18 |
| 35 | South Carolina | 12.56 |
| 36 | Louisiana | 12.77 |
| 37 | South Dakota | 12.89 |
| 38 | Missouri | 13.19 |
| 39 | New York | 13.32 |
| 40 | Vermont | 13.63 |
| 41 | Mississippi | 14.23 |
| 42 | Hawaii | 14.28 |
| 43 | Alabama | 14.38 |
| 44 | North Dakota | 14.98 |
| 45 | Oklahoma | 15.00 |
| 46 | Wyoming | 15.36 |
| 47 | Arkansas | 16.01 |
| 48 | Montana | 19.30 |
| 49 | New Mexico | 19.68 |
| 50 | West Virginia | 25.58 |
| 51 | Alaska | 26.55 |

*States are sorted and colored by the risk of error implied by the process statistic—from lowest risk (rank 1, lightest color) to highest risk (rank 51, darkest color). The bars facilitate comparisons between states, having 0 length for the lowest risk state and filling the entire cell for the highest.

### Process Statistic 2: Questionnaires Without Identification not on MAF

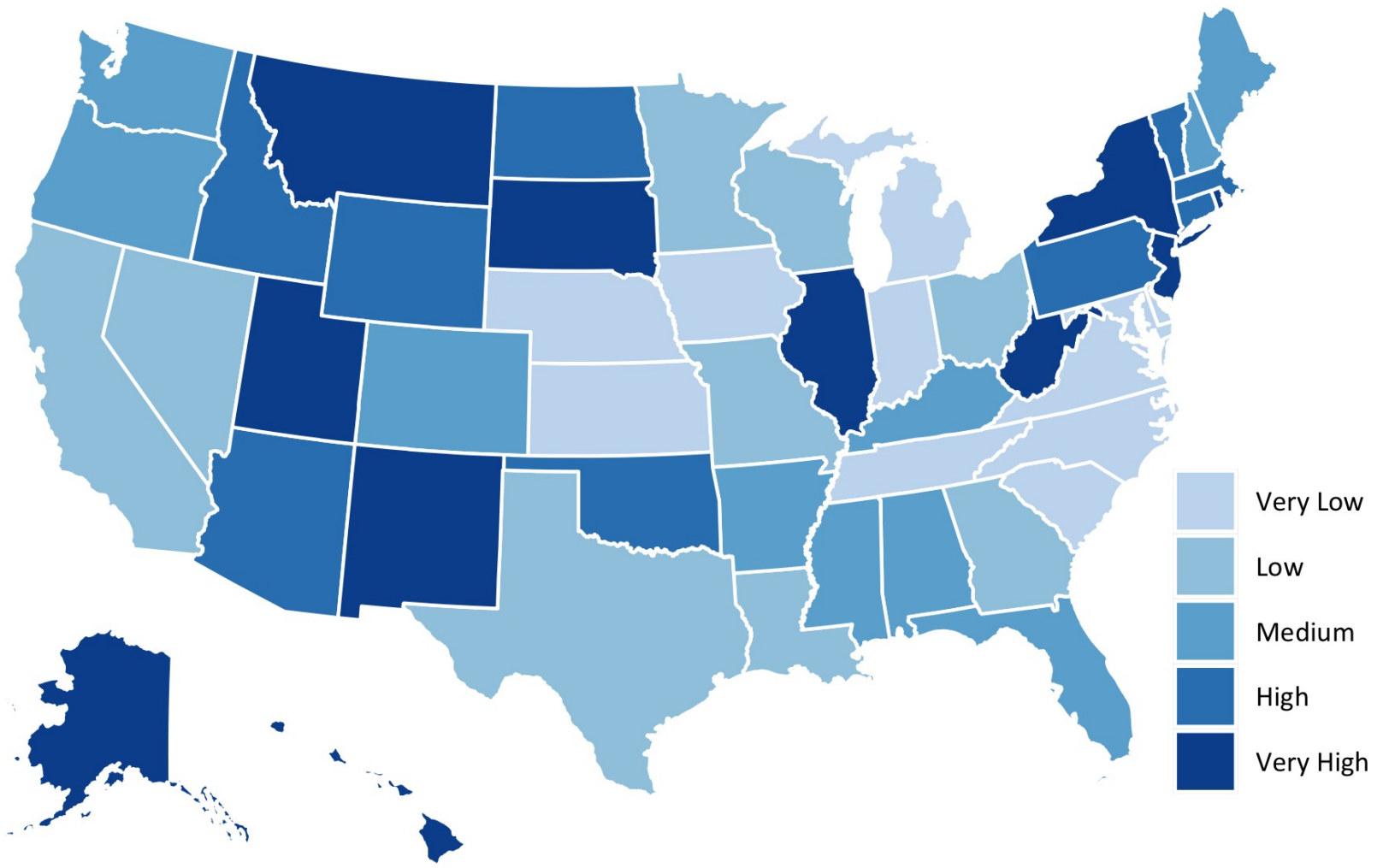

**How is this process statistic calculated?**
Percent of HUs submitting questionnaires without census IDs and no matching address was found on the MAF for 2020.

**How is this process statistic interpreted?**
Ideally, every response includes a census ID, linking it to the Census Bureau address list. Some responses lack IDs, and the Bureau must carefully examine these responses to ensure that they come from valid addresses. In particular, the Bureau checks if the address is in the MAF. But nonstandard addresses (e.g., those without a number or street), may not be in the MAF. This statistic reflects quality because responses without ID that are not in the MAF are at higher risk of being incorrectly included or excluded in the census count.

**Key Points**

- Nationwide, nearly 9% of households without census IDs had no match in the MAF.
- The percentages varied widely by state, from 4.3% (MD) to 21.6% (AK).
- Nonstandard addresses in urban and rural areas are a contributing factor.

**Future Research**

- Does an increase in non-ID returns decrease count accuracy?
- Did encouraging non-ID responses also encourage the submission of addresses that could not be matched to the MAF?
- Did residence issues related to the pandemic affect the evaluation of nonmatches in the field?

| Rank | State | Process Statistic 2: Percent of HUs submitting questionnaires without census IDs and no matching address was found on the MAF for 2020* |
|---|---|---|
| 1 | Maryland | 4.28 |
| 2 | Virginia | 4.80 |
| 3 | South Carolina | 4.90 |
| 4 | Kansas | 4.95 |
| 5 | Iowa | 5.14 |
| 6 | Tennessee | 5.15 |
| 7 | North Carolina | 5.29 |
| 8 | Delaware | 5.38 |
| 9 | Michigan | 5.50 |
| 10 | Indiana | 5.51 |
| 11 | Nebraska | 5.85 |
| 12 | Georgia | 5.88 |
| 13 | Minnesota | 6.08 |
| 14 | Ohio | 6.09 |
| 15 | Louisiana | 6.45 |
| 16 | Missouri | 6.46 |
| 17 | California | 6.47 |
| 18 | Nevada | 6.63 |
| 19 | Wisconsin | 6.81 |
| 20 | Texas | 6.97 |
| 21 | Washington, D.C. | 7.16 |
| 22 | Alabama | 7.21 |
| 23 | Mississippi | 7.25 |
| 24 | New Hampshire | 7.45 |
| 25 | Arkansas | 7.52 |
| 26 | Oregon | 7.53 |
| 27 | Maine | 7.79 |
| 28 | Florida | 7.90 |
| 29 | Kentucky | 7.93 |
| 30 | Colorado | 8.16 |
| 31 | Washington | 8.37 |
| 32 | Idaho | 8.53 |
| 33 | Pennsylvania | 8.76 |
| | United States | 8.79 |
| 34 | Vermont | 9.24 |
| 35 | Oklahoma | 9.71 |
| 36 | Massachusetts | 10.60 |
| 37 | Wyoming | 10.68 |
| 38 | North Dakota | 11.02 |
| 39 | Arizona | 11.25 |
| 40 | Connecticut | 11.58 |
| 41 | Illinois | 12.03 |
| 42 | Montana | 12.30 |
| 43 | New Jersey | 12.71 |
| 44 | Utah | 13.59 |
| 45 | Hawaii | 14.09 |
| 46 | West Virginia | 15.07 |
| 47 | Rhode Island | 15.49 |
| 48 | New York | 15.76 |
| 49 | New Mexico | 16.42 |
| 50 | South Dakota | 19.20 |
| 51 | Alaska | 21.55 |

*States are sorted and colored by the risk of error implied by the process statistic—from lowest risk (rank 1, lightest color) to highest risk (rank 51, darkest color). The bars facilitate comparisons between states, having 0 length for the lowest risk state and filling the entire cell for the highest.

### Process Statistic 3: Multiple Responses

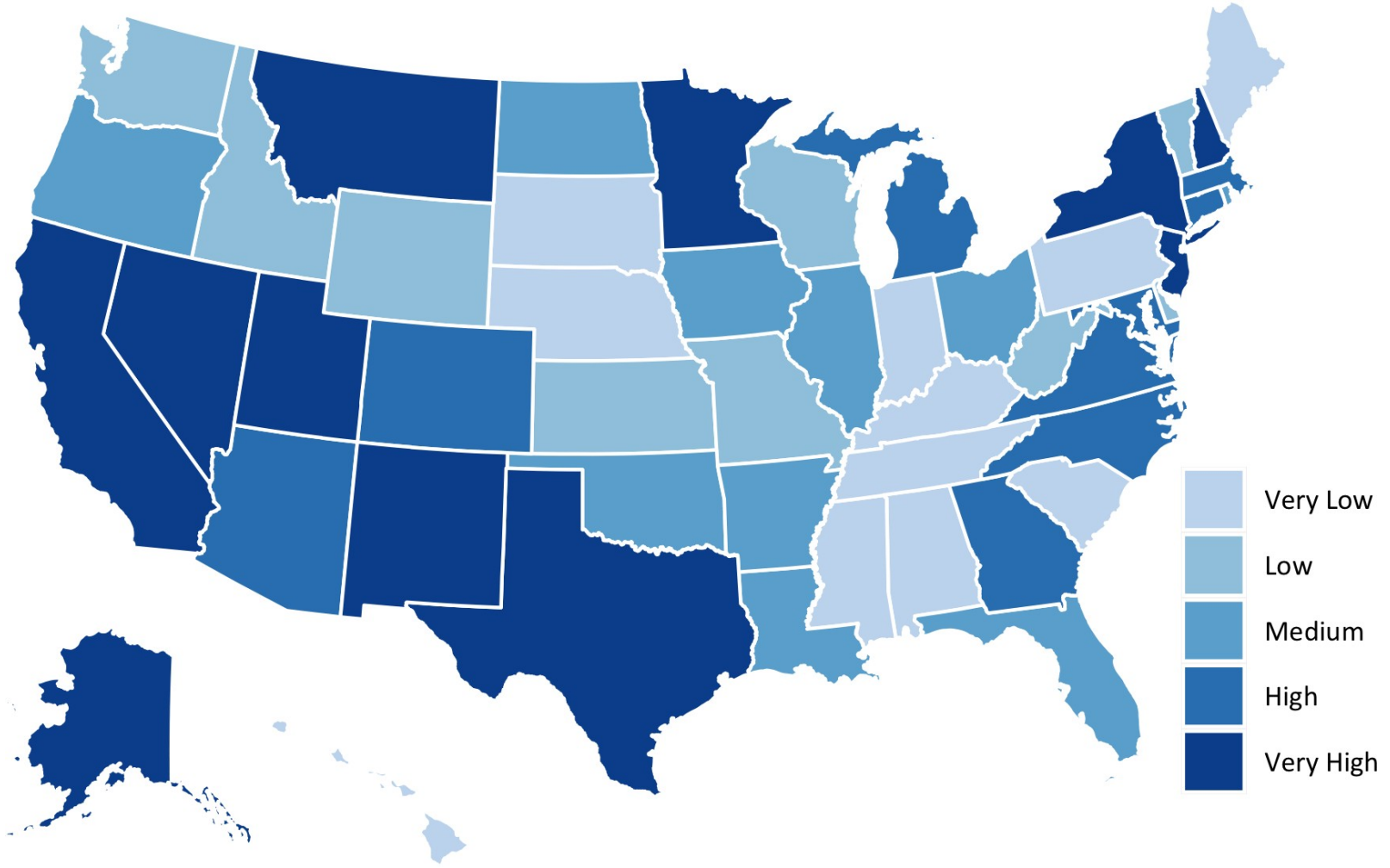

**How is this process statistic calculated?**
Percent of occupied HUs with two or more responses from various sources for 2020 minus the corresponding percentage for 2010.

**How is this process statistic interpreted?**
Ideally, one member of every household fills out one and only questionnaire. But inevitably some households return multiple questionnaires. Other operations also produce multiple returns. This statistic reflects quality because the Census Bureau must correctly identify and eliminate multiple returns and that process, especially for duplicate questionnaires, can increase error risk. Little is known, however, about the overall accuracy of deduplication methods.

**Key Points**
- All states had double-digit increases in the percentage of multiple responses in 2020. Nationwide, the percentage was 18%.
- The increase in the percentage of multiple response households ranged from 15% (ME) to 20% (NM).

**Future Research**
- How do the various types of multiple responses contribute to census error?
- What are the characteristics of households with multiple responses?
- What process is used for choosing between multiple responses received from the same household?

| Rank | State | Percent of occupied HUs with two or more responses from various sources for 2020 | Process Statistic 3: Percent of occupied HUs with two or more responses from various sources for 2020 minus the corresponding percentage for 2010* |
|---|---|---|---|
| 1 | Maine | 22.87 | 14.98 |
| 2 | Alabama | 25.44 | 15.18 |
| 3 | Mississippi | 25.70 | 15.25 |
| 4 | South Carolina | 25.59 | 15.38 |
| 5 | Indiana | 22.92 | 15.43 |
| 6 | Kentucky | 23.58 | 15.45 |
| 7 | Hawaii | 29.96 | 15.66 |
| 8 | Tennessee | 24.26 | 15.70 |
| 9 | Nebraska | 22.49 | 15.73 |
| 10 | Pennsylvania | 24.18 | 15.90 |
| 11 | South Dakota | 22.92 | 15.94 |
| 12 | Idaho | 24.60 | 15.95 |
| 13 | Kansas | 23.56 | 16.04 |
| 14 | Missouri | 23.76 | 16.24 |
| 15 | Wisconsin | 22.72 | 16.31 |
| 16 | Washington, D.C. | 29.06 | 16.52 |
| 17 | West Virginia | 24.96 | 16.52 |
| 18 | Wyoming | 24.83 | 16.55 |
| 19 | Washington | 25.41 | 16.60 |
| 20 | Vermont | 24.87 | 16.67 |
| 21 | Delaware | 25.65 | 16.68 |
| 22 | Arkansas | 24.81 | 16.69 |
| 23 | North Dakota | 23.13 | 16.76 |
| 24 | Oklahoma | 25.66 | 16.91 |
| 25 | Florida | 26.39 | 16.97 |
| 26 | Ohio | 24.32 | 16.97 |
| 27 | Illinois | 25.50 | 17.04 |
| 28 | Rhode Island | 25.57 | 17.10 |
| 29 | Louisiana | 26.84 | 17.20 |
| 30 | Iowa | 23.43 | 17.21 |
| 31 | Oregon | 25.33 | 17.23 |
| 32 | North Carolina | 26.85 | 17.35 |
| 33 | Michigan | 24.98 | 17.39 |
| 34 | Connecticut | 26.15 | 17.41 |
| 35 | Massachusetts | 26.71 | 17.46 |
| | United States | 26.47 | 17.53 |
| 36 | Virginia | 26.58 | 17.65 |
| 37 | Maryland | 27.31 | 17.78 |
| 38 | Georgia | 28.15 | 17.88 |
| 39 | Colorado | 25.83 | 18.09 |
| 40 | Arizona | 27.39 | 18.13 |
| 41 | Utah | 27.90 | 18.32 |
| 42 | Minnesota | 25.17 | 18.34 |
| 43 | New Jersey | 28.18 | 18.52 |
| 44 | California | 28.56 | 18.82 |
| 45 | New Hampshire | 26.56 | 18.87 |
| 46 | Nevada | 27.57 | 18.97 |
| 47 | Texas | 28.64 | 19.04 |
| 48 | New York | 29.86 | 19.61 |
| 49 | Montana | 26.54 | 19.88 |
| 50 | Alaska | 32.55 | 20.14 |
| 51 | New Mexico | 29.31 | 20.15 |

*States are sorted and colored by the risk of error implied by the process statistic—from lowest risk (rank 1, lightest color) to highest risk (rank 51, darkest color). The bars facilitate comparisons between states, having 0 length for the lowest risk state and filling the entire cell for the highest.

### Process Statistic 4: Usual Residence at College

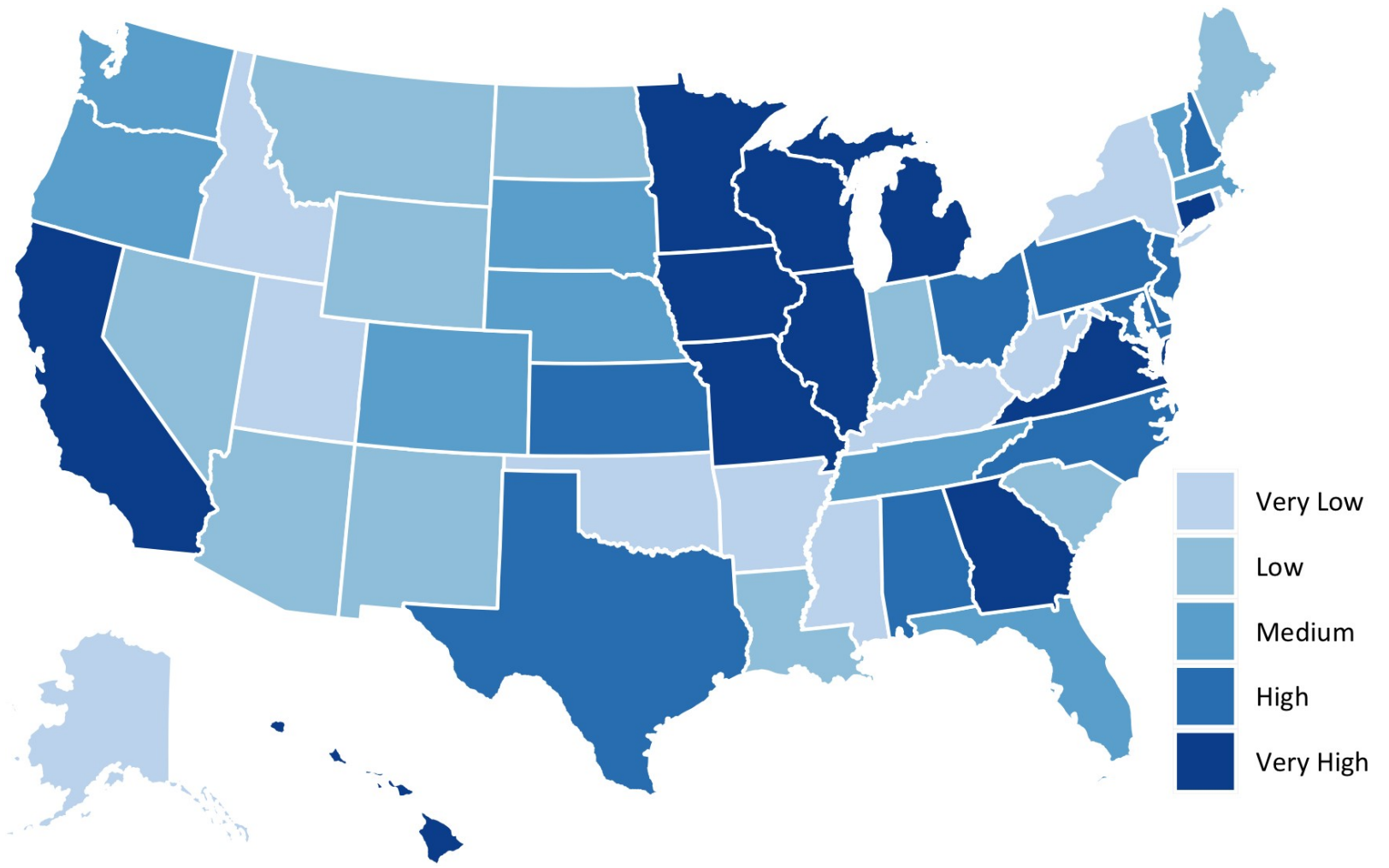

**How is this process statistic calculated?**
Percent of occupied HUs with two or more people where one or more occupant indicated their usual residence was at college for 2020 minus the corresponding percentage for 2010.

**How is this process statistic interpreted?**
Ideally, students in college residences as of April 1, 2020, are counted as living in college residences. But because of the pandemic, many students moved and instead responded as living in a household away from college. This PS reflects quality because the Census Bureau must decide whether to reassign these students, increasing the risk of error. Little is known, however, about the overall accuracy of reassignment.

**Key Points**

- The percent of households with a college student reporting a Usual Residence at College (URC) rose modestly between 2010 and 2020, roughly half a percent nationwide.
- The increase in households with a college student reporting a URC ranged from -0.11% (DC) to 0.89% (VA).

**Future Research**

- Does the reassignment of URC residents increase coverage error?
- How often does the Bureau fail to correctly reassign URCs?
- How much does a failure to correctly reassign URCs contribute to coverage error?

| Rank | State | Percent of occupied HUs with 2+ persons where 1+ indicated usual residence was college for 2020 | Process Statistic 4: Percent of occupied HUs with two or more people where one or more occupant indicated their usual residence was at college for 2020 minus the corresponding percentage for 2010* |
|---|---|---|---|
| 1 | Washington, D.C. | 1.40 | -0.11 |
| 2 | West Virginia | 0.98 | 0.13 |
| 3 | Idaho | 0.95 | 0.13 |
| 4 | Utah | 1.29 | 0.17 |
| 5 | Rhode Island | 1.12 | 0.17 |
| 6 | Alaska | 0.84 | 0.18 |
| 7 | Arkansas | 0.90 | 0.20 |
| 8 | Mississippi | 1.44 | 0.20 |
| 9 | New York | 1.46 | 0.21 |
| 10 | Oklahoma | 0.93 | 0.23 |
| 11 | Kentucky | 1.00 | 0.23 |
| 12 | South Carolina | 0.74 | 0.25 |
| 13 | Nevada | 1.20 | 0.25 |
| 14 | Louisiana | 1.26 | 0.28 |
| 15 | Wyoming | 1.09 | 0.31 |
| 16 | Maine | 1.15 | 0.32 |
| 17 | Arizona | 0.95 | 0.35 |
| 18 | Indiana | 1.17 | 0.37 |
| 19 | Montana | 1.06 | 0.37 |
| 20 | North Dakota | 1.05 | 0.39 |
| 21 | New Mexico | 1.18 | 0.39 |
| 22 | Vermont | 1.41 | 0.40 |
| 23 | Florida | 1.19 | 0.42 |
| 24 | Tennessee | 1.21 | 0.43 |
| 25 | Oregon | 1.08 | 0.44 |
| 26 | Massachusetts | 1.81 | 0.45 |
| 27 | Washington | 1.15 | 0.47 |
| 28 | South Dakota | 1.37 | 0.51 |
| 29 | Nebraska | 1.32 | 0.52 |
| | United States | 1.45 | 0.53 |
| 30 | Colorado | 1.24 | 0.54 |
| 31 | New Hampshire | 1.74 | 0.55 |
| 32 | Maryland | 1.64 | 0.55 |
| 33 | Ohio | 1.40 | 0.56 |
| 34 | North Carolina | 1.93 | 0.58 |
| 35 | New Jersey | 1.46 | 0.58 |
| 36 | Texas | 1.34 | 0.59 |
| 37 | Kansas | 1.46 | 0.59 |
| 38 | Alabama | 1.51 | 0.60 |
| 39 | Delaware | 1.54 | 0.62 |
| 40 | Pennsylvania | 1.73 | 0.62 |
| 41 | Missouri | 1.71 | 0.64 |
| 42 | Illinois | 1.38 | 0.64 |
| 43 | Connecticut | 1.87 | 0.66 |
| 44 | Hawaii | 1.36 | 0.67 |
| 45 | Georgia | 1.76 | 0.68 |
| 46 | Iowa | 1.53 | 0.76 |
| 47 | Michigan | 1.68 | 0.76 |
| 48 | Wisconsin | 1.51 | 0.77 |
| 49 | California | 1.62 | 0.78 |
| 50 | Minnesota | 1.67 | 0.79 |
| 51 | Virginia | 1.82 | 0.89 |

*States are sorted and colored by the risk of error implied by the process statistic—from lowest risk (rank 1, lightest color) to highest risk (rank 51, darkest color). The bars facilitate comparisons between states, having 0 length for the lowest risk state and filling the entire cell for the highest.

### Process Statistic 5: Responses Obtained by Proxy

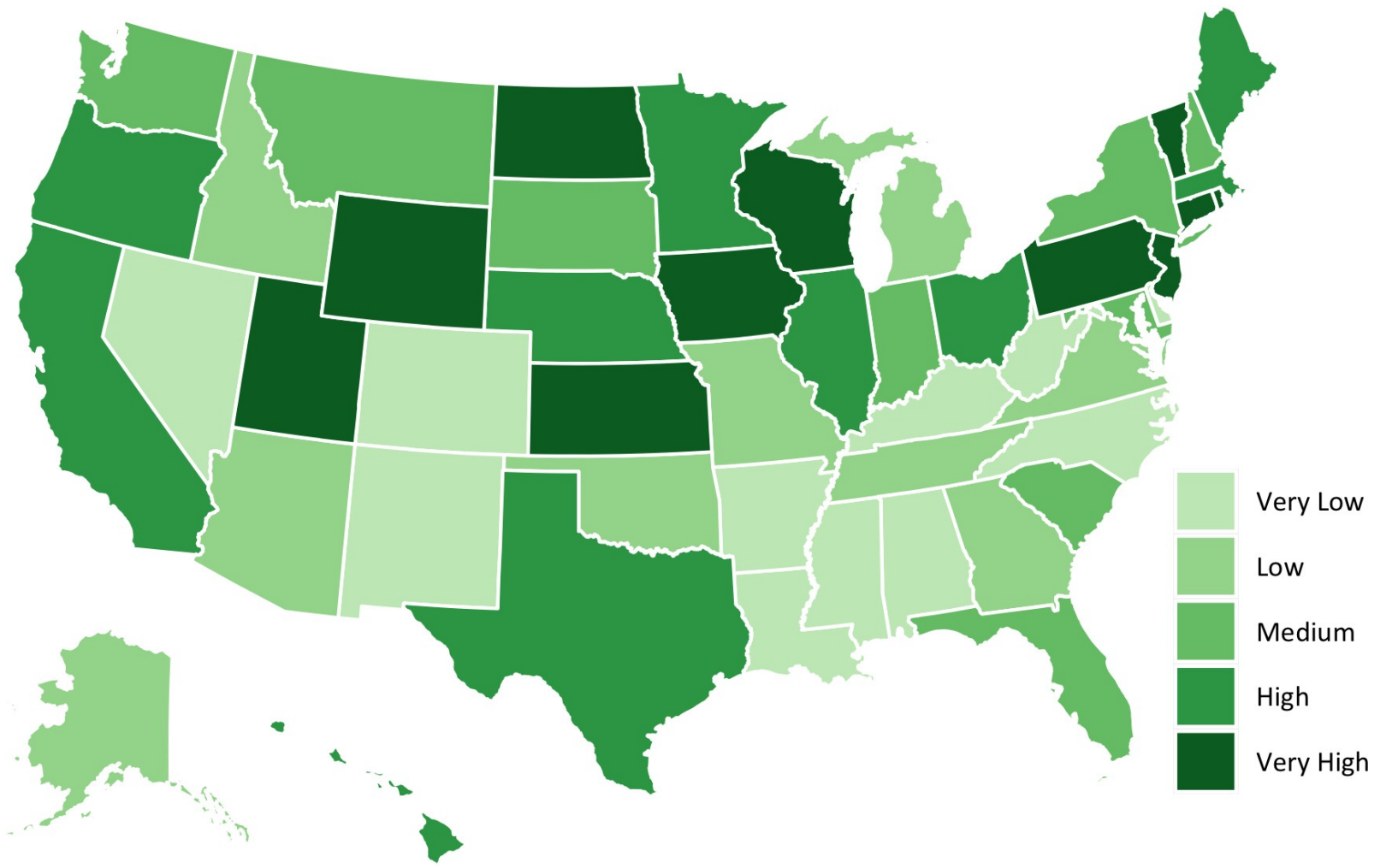

**How is this process statistic calculated?**
Percent of persons in occupied HUs whose count was obtained by proxy interview (during NRFU) for 2020 minus the corresponding percentage for 2010.

**How is this process statistic interpreted?**
Ideally, the Census Bureau interviews a member of every household that did not self-respond. But inevitably some households cannot be reached and enumerators must interview proxies, such as a neighbor or building manager. This PS reflects quality because research from prior censuses has demonstrated that proxy interviews are more likely to contain errors.

**Key Points**

- Nationwide, the percentage of proxies declined 0.35% in 2020 relative to 2010. The greatest declines were in WV, LA, and MS.
- Four states—KS, VT, UT, and RI—saw increases in the percentage of proxies by 0.5% or more. RI increased by 1.7%.

**Future Research**

- How complete were proxy responses?
- Did the use of administrative records lead to a reduction in the use of proxies in the 2020 census data collection?

| Rank | State | Percent of persons in occupied HUs whose count was obtained by proxy interview for 2020 | Process Statistic 5: Percent of persons in occupied HUs whose count was obtained by proxy interview for 2020 minus the corresponding percentage for 2010* |
|---|---|---|---|
| 1 | West Virginia | 4.44 | -2.11 |
| 2 | Louisiana | 4.56 | -1.90 |
| 3 | Mississippi | 3.46 | -1.89 |
| 4 | Delaware | 4.29 | -1.78 |
| 5 | Kentucky | 3.94 | -1.74 |
| 6 | Nevada | 5.69 | -1.73 |
| 7 | Alabama | 4.34 | -1.72 |
| 8 | New Mexico | 5.47 | -1.70 |
| 9 | Arkansas | 3.99 | -1.37 |
| 10 | North Carolina | 4.71 | -1.17 |
| 11 | Colorado | 4.71 | -1.11 |
| 12 | Tennessee | 4.44 | -1.07 |
| 13 | Washington, D.C. | 6.34 | -0.98 |
| 14 | Missouri | 3.74 | -0.91 |
| 15 | Michigan | 3.43 | -0.87 |
| 16 | Georgia | 5.28 | -0.87 |
| 17 | Arizona | 5.87 | -0.79 |
| 18 | Idaho | 3.79 | -0.64 |
| 19 | Alaska | 5.52 | -0.62 |
| 20 | Virginia | 4.17 | -0.49 |
| 21 | Oklahoma | 5.23 | -0.42 |
| 22 | Montana | 4.62 | -0.42 |
| 23 | Washington | 4.04 | -0.38 |
| 24 | South Dakota | 4.17 | -0.37 |
| 25 | New Hampshire | 3.70 | -0.35 |
| | United States | 4.65 | -0.35 |
| 26 | Maryland | 4.30 | -0.31 |
| 27 | New York | 5.44 | -0.29 |
| 28 | South Carolina | 4.79 | -0.28 |
| 29 | Florida | 5.17 | -0.28 |
| 30 | Indiana | 4.04 | -0.21 |
| 31 | Oregon | 4.36 | -0.11 |
| 32 | California | 4.42 | -0.11 |
| 33 | Nebraska | 3.89 | -0.07 |
| 34 | Ohio | 4.33 | -0.07 |
| 35 | Minnesota | 3.40 | -0.03 |
| 36 | Hawaii | 5.66 | -0.01 |
| 37 | Massachusetts | 4.40 | 0.00 |
| 38 | Maine | 4.01 | 0.03 |
| 39 | Texas | 5.84 | 0.07 |
| 40 | Illinois | 4.56 | 0.09 |
| 41 | New Jersey | 4.59 | 0.14 |
| 42 | Connecticut | 4.48 | 0.21 |
| 43 | Wisconsin | 3.64 | 0.21 |
| 44 | Wyoming | 5.16 | 0.24 |
| 45 | North Dakota | 4.40 | 0.28 |
| 46 | Iowa | 3.95 | 0.29 |
| 47 | Pennsylvania | 4.44 | 0.34 |
| 48 | Kansas | 4.30 | 0.69 |
| 49 | Vermont | 4.08 | 0.69 |
| 50 | Utah | 4.35 | 0.77 |
| 51 | Rhode Island | 6.65 | 1.74 |

*States are sorted and colored by the risk of error implied by the process statistic—from lowest risk (rank 1, lightest color) to highest risk (rank 51, darkest color). The bars facilitate comparisons between states, having 0 length for the lowest risk state and filling the entire cell for the highest.

### Process Statistic 6: Enumerations With Only a Population Count

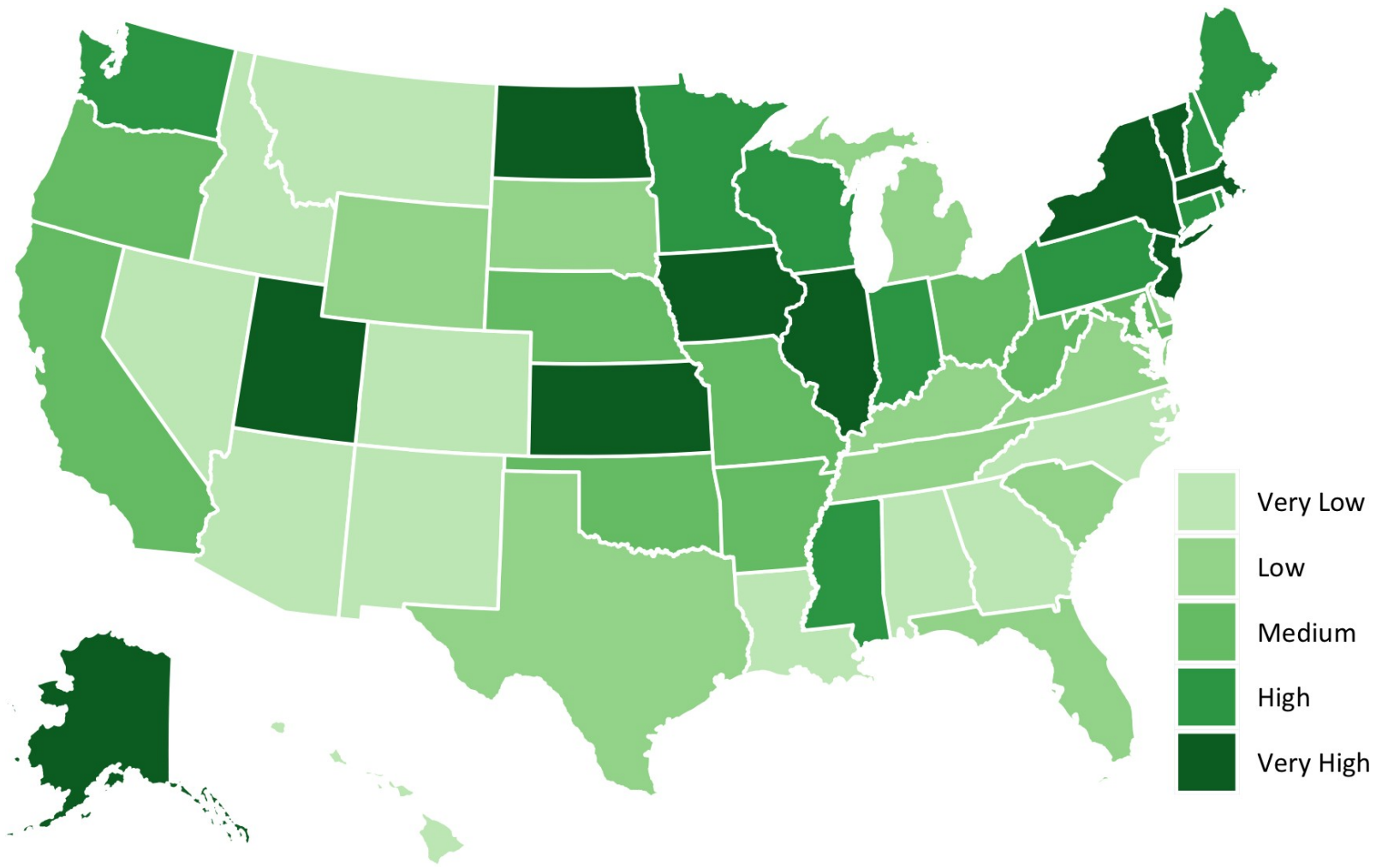

**How is this process statistic calculated?**
Percent of occupied HUs where only a population count was obtained for 2020 minus the corresponding percentage for 2010.

**How is this process statistic interpreted?**
Ideally, every response contains complete information on all residents. But inevitably some responses are population count only, meaning only the number of persons in each household is known. These responses can arise from interviews with apprehensive household members, which may have been more common during the pandemic. This PS reflects quality because incomplete responses are more likely to have incorrect information.

**Key Points**

- Nationwide, the percent of population count–only households increased roughly half a percentage point between 2010 and 2020.
- The change in percentages varied widely by state, from increasing 1.6% (NY) to declining 1.3% (HI).
- The increase in 13 states exceeded 1%, while 3 states declined more than 1%.

**Future Research**

- How is the timestamp of an enumeration related to the frequency of population count–only households?
- How much did proxy interviews and administrative records contribute to the number of population count–only households?

| Rank | State | Percent of occupied HUs where only a population count was obtained for 2020 | Process Statistic 6: Percent of occupied HUs where only a population count was obtained for 2020 minus the corresponding percentage for 2010* |
|---|---|---|---|
| 1 | Hawaii | 1.69 | -1.25 |
| 2 | Nevada | 2.65 | -1.22 |
| 3 | Arizona | 2.10 | -1.02 |
| 4 | Colorado | 2.18 | -0.84 |
| 5 | New Mexico | 2.51 | -0.62 |
| 6 | Georgia | 2.11 | -0.61 |
| 7 | Montana | 2.03 | -0.60 |
| 8 | Idaho | 1.89 | -0.38 |
| 9 | Alabama | 2.01 | -0.26 |
| 10 | Louisiana | 2.27 | -0.19 |
| 11 | North Carolina | 2.19 | -0.11 |
| 12 | Delaware | 2.38 | -0.08 |
| 13 | Wyoming | 2.25 | 0.04 |
| 14 | Texas | 2.34 | 0.06 |
| 15 | Virginia | 2.01 | 0.13 |
| 16 | Kentucky | 1.87 | 0.13 |
| 17 | Florida | 2.46 | 0.15 |
| 18 | South Dakota | 1.77 | 0.29 |
| 19 | South Carolina | 1.94 | 0.30 |
| 20 | Tennessee | 2.12 | 0.35 |
| 21 | Michigan | 2.08 | 0.51 |
| | United States | 2.28 | 0.53 |
| 22 | Arkansas | 1.89 | 0.54 |
| 23 | West Virginia | 1.69 | 0.62 |
| 24 | Missouri | 2.26 | 0.62 |
| 25 | Oklahoma | 1.87 | 0.63 |
| 26 | Nebraska | 1.95 | 0.68 |
| 27 | Ohio | 2.08 | 0.70 |
| 28 | California | 2.44 | 0.79 |
| 29 | Maryland | 2.44 | 0.79 |
| 30 | Oregon | 2.00 | 0.80 |
| 31 | Rhode Island | 2.46 | 0.80 |
| 32 | Maine | 1.85 | 0.85 |
| 33 | Mississippi | 2.09 | 0.87 |
| 34 | New Hampshire | 1.91 | 0.88 |
| 35 | Washington | 2.23 | 0.88 |
| 36 | Indiana | 2.04 | 0.91 |
| 37 | Pennsylvania | 2.03 | 0.93 |
| 38 | Minnesota | 1.83 | 0.95 |
| 39 | Wisconsin | 2.04 | 1.01 |
| 40 | Connecticut | 2.35 | 1.06 |
| 41 | Iowa | 1.87 | 1.09 |
| 42 | Illinois | 2.69 | 1.10 |
| 43 | Utah | 2.52 | 1.13 |
| 44 | North Dakota | 2.03 | 1.15 |
| 45 | Massachusetts | 2.27 | 1.18 |
| 46 | Kansas | 1.85 | 1.20 |
| 47 | Washington, D.C. | 3.31 | 1.26 |
| 48 | Alaska | 2.29 | 1.30 |
| 49 | New Jersey | 2.60 | 1.37 |
| 50 | Vermont | 1.79 | 1.41 |
| 51 | New York | 3.17 | 1.59 |

*States are sorted and colored by the risk of error implied by the process statistic—from lowest risk (rank 1, lightest color) to highest risk (rank 51, darkest color). The bars facilitate comparisons between states, having 0 length for the lowest risk state and filling the entire cell for the highest.

### Process Statistic 7: Enumerations via Administrative Records

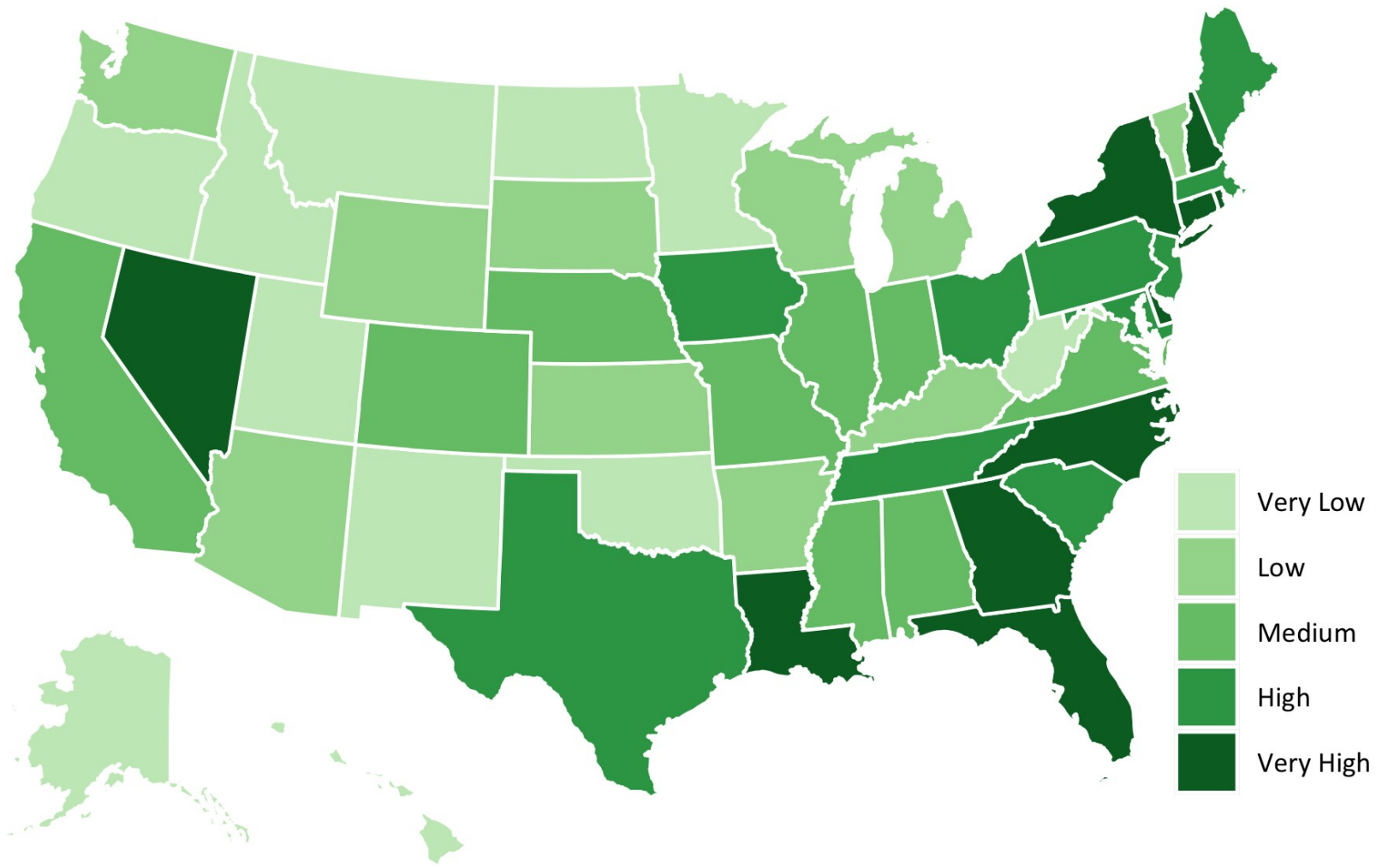

**How is this process statistic calculated?**
Percent of occupied HUs enumerated by administrative records for 2020.

**How is this process statistic interpreted?**
Ideally, the Census Bureau interviews a member of every household that did not self-respond. But inevitably some households cannot be reached. Administrative records, new for 2020, can provide the missing information—both housing status (occupied, vacant, etc.) and characteristics (race, gender, etc.). This PS reflects quality because administrative records may be outdated, inaccurate, or incomplete, increasing the risk of error.

**Key Points**

- Nationwide, the percentage of households enumerated with administrative records was 3.8%.
- Percentages varied widely by state, from 1.7% (HI) to more than 5% (RI and LA).
- Administrative records may have reduced reliance on less accurate enumeration methods, such as unknowledgeable proxies, and imputation.

**Future Research**

- Did using administrative records increase coverage error?
- How does the error rate compare with alternatives such as using proxies and statistical imputation?

| Rank | State | Process Statistic 7: Percent of occupied HUs enumerated by administrative records for 2020* |
|---|---|---|
| 1 | Hawaii | 1.65 |
| 2 | Alaska | 2.11 |
| 3 | Idaho | 2.51 |
| 4 | Utah | 2.54 |
| 5 | New Mexico | 2.68 |
| 6 | Montana | 2.73 |
| 7 | West Virginia | 2.74 |
| 8 | Minnesota | 2.85 |
| 9 | Oklahoma | 2.85 |
| 10 | North Dakota | 2.90 |
| 11 | Oregon | 2.97 |
| 12 | Kansas | 3.05 |
| 13 | Vermont | 3.09 |
| 14 | South Dakota | 3.10 |
| 15 | Wisconsin | 3.17 |
| 16 | Kentucky | 3.18 |
| 17 | Washington | 3.25 |
| 18 | Arkansas | 3.36 |
| 19 | Michigan | 3.36 |
| 20 | Wyoming | 3.38 |
| 21 | Arizona | 3.41 |
| 22 | Nebraska | 3.47 |
| 23 | Alabama | 3.51 |
| 24 | Colorado | 3.55 |
| 25 | Missouri | 3.61 |
| 26 | Indiana | 3.62 |
| 27 | Virginia | 3.64 |
| 28 | Illinois | 3.64 |
| 29 | California | 3.75 |
| | United States | 3.84 |
| 30 | Mississippi | 3.89 |
| 31 | Tennessee | 3.94 |
| 32 | Texas | 3.97 |
| 33 | Pennsylvania | 4.04 |
| 34 | Maryland | 4.04 |
| 35 | Ohio | 4.07 |
| 36 | New Jersey | 4.09 |
| 37 | Maine | 4.12 |
| 38 | Massachusetts | 4.19 |
| 39 | Iowa | 4.23 |
| 40 | South Carolina | 4.26 |
| 41 | Connecticut | 4.27 |
| 42 | Georgia | 4.32 |
| 43 | New York | 4.37 |
| 44 | Nevada | 4.39 |
| 45 | Florida | 4.43 |
| 46 | New Hampshire | 4.46 |
| 47 | North Carolina | 4.52 |
| 48 | Delaware | 4.87 |
| 49 | Washington, D.C. | 4.98 |
| 50 | Rhode Island | 5.09 |
| 51 | Louisiana | 5.12 |

*States are sorted and colored by the risk of error implied by the process statistic—from lowest risk (rank 1, lightest color) to highest risk (rank 51, darkest color). The bars facilitate comparisons between states, having 0 length for the lowest risk state and filling the entire cell for the highest.

### Process Statistic 8: MAF Addresses Having Imputed Status

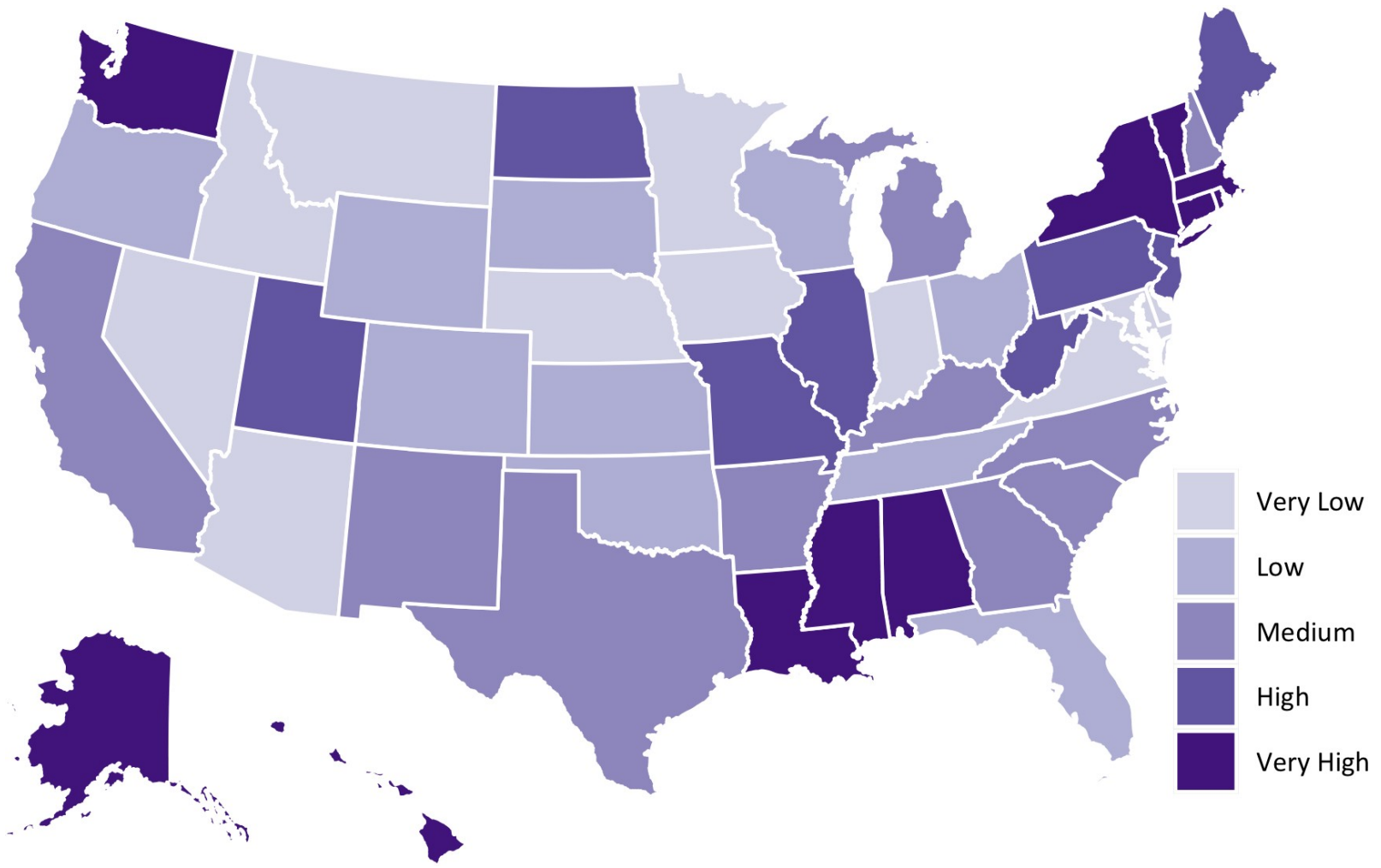

**How is this process statistic calculated?**
Percent of MAF units whose status was imputed for 2020 minus the corresponding percentage for 2010.

**How is this process statistic interpreted?**
Ideally, the Census Bureau can determine whether a HU is occupied after several visits or with administrative records. But inevitably, these procedures may be inconclusive. The Bureau may then use information on neighboring HUs to predict the status, a process known as imputation. This PS measures quality because imputations may incorrectly classify vacant or nonexistent units as occupied or occupied units as vacant, increasing the risk of error.

**Key Points**

- All states experienced an increase in the percentage of HUs whose status was imputed. Nationwide, the increase was roughly three-quarters of a percentage point.
- In five states—LA, NY, MA, RI, and HI—the percentages increased by one point or more.
- The pandemic may have decreased the accuracy of imputations.

**Future Research**

- Do status imputations increase coverage errors?
- How much did the pandemic affect status imputations?

| Rank | State | Percent of MAF units whose status was imputed for 2020 | Process Statistic 8: Percent of MAF units whose status was imputed for 2020 minus the corresponding percentage for 2010* |
|---|---|---|---|
| 1 | Delaware | 0.64 | 0.25 |
| 2 | Arizona | 0.68 | 0.48 |
| 3 | Idaho | 0.71 | 0.53 |
| 4 | Iowa | 0.61 | 0.54 |
| 5 | Nevada | 0.78 | 0.55 |
| 6 | Maryland | 0.71 | 0.55 |
| 7 | Montana | 0.82 | 0.56 |
| 8 | Minnesota | 0.68 | 0.60 |
| 9 | Virginia | 0.72 | 0.60 |
| 10 | Indiana | 0.70 | 0.60 |
| 11 | Nebraska | 0.71 | 0.62 |
| 12 | Oregon | 0.74 | 0.63 |
| 13 | South Dakota | 0.79 | 0.63 |
| 14 | Florida | 0.81 | 0.64 |
| 15 | Ohio | 0.73 | 0.65 |
| 16 | Oklahoma | 0.82 | 0.65 |
| 17 | Kansas | 0.74 | 0.66 |
| 18 | Colorado | 0.77 | 0.66 |
| 19 | Tennessee | 0.80 | 0.67 |
| 20 | Wyoming | 0.89 | 0.67 |
| 21 | Wisconsin | 0.78 | 0.67 |
| 22 | North Carolina | 0.89 | 0.67 |
| 23 | Kentucky | 0.90 | 0.69 |
| 24 | New Hampshire | 0.80 | 0.69 |
| 25 | South Carolina | 0.91 | 0.69 |
| 26 | Michigan | 0.81 | 0.69 |
| 27 | Texas | 0.85 | 0.70 |
| 28 | Arkansas | 0.91 | 0.71 |
| 29 | New Mexico | 0.99 | 0.71 |
| 30 | Georgia | 0.90 | 0.71 |
| 31 | California | 0.81 | 0.72 |
| 32 | Illinois | 0.85 | 0.73 |
| 33 | Pennsylvania | 0.86 | 0.73 |
|  | United States | 0.88 | 0.74 |
| 34 | Washington, D.C. | 1.01 | 0.74 |
| 35 | Missouri | 0.86 | 0.75 |
| 36 | West Virginia | 0.99 | 0.75 |
| 37 | North Dakota | 0.89 | 0.77 |
| 38 | New Jersey | 0.92 | 0.78 |
| 39 | Utah | 0.90 | 0.80 |
| 40 | Maine | 0.93 | 0.80 |
| 41 | Alaska | 1.01 | 0.81 |
| 42 | Washington | 0.93 | 0.81 |
| 43 | Connecticut | 0.98 | 0.84 |
| 44 | Mississippi | 1.04 | 0.84 |
| 45 | Vermont | 0.99 | 0.85 |
| 46 | Alabama | 1.09 | 0.90 |
| 47 | Hawaii | 1.16 | 1.00 |
| 48 | Rhode Island | 1.15 | 1.03 |
| 49 | Massachusetts | 1.17 | 1.06 |
| 50 | New York | 1.39 | 1.19 |
| 51 | Louisiana | 1.53 | 1.36 |

*States are sorted and colored by the risk of error implied by the process statistic—from lowest risk (rank 1, lightest color) to highest risk (rank 51, darkest color). The bars facilitate comparisons between states, having 0 length for the lowest risk state and filling the entire cell for the highest.

### Process Statistic 9: Occupied Housing Units With Imputed Population Counts

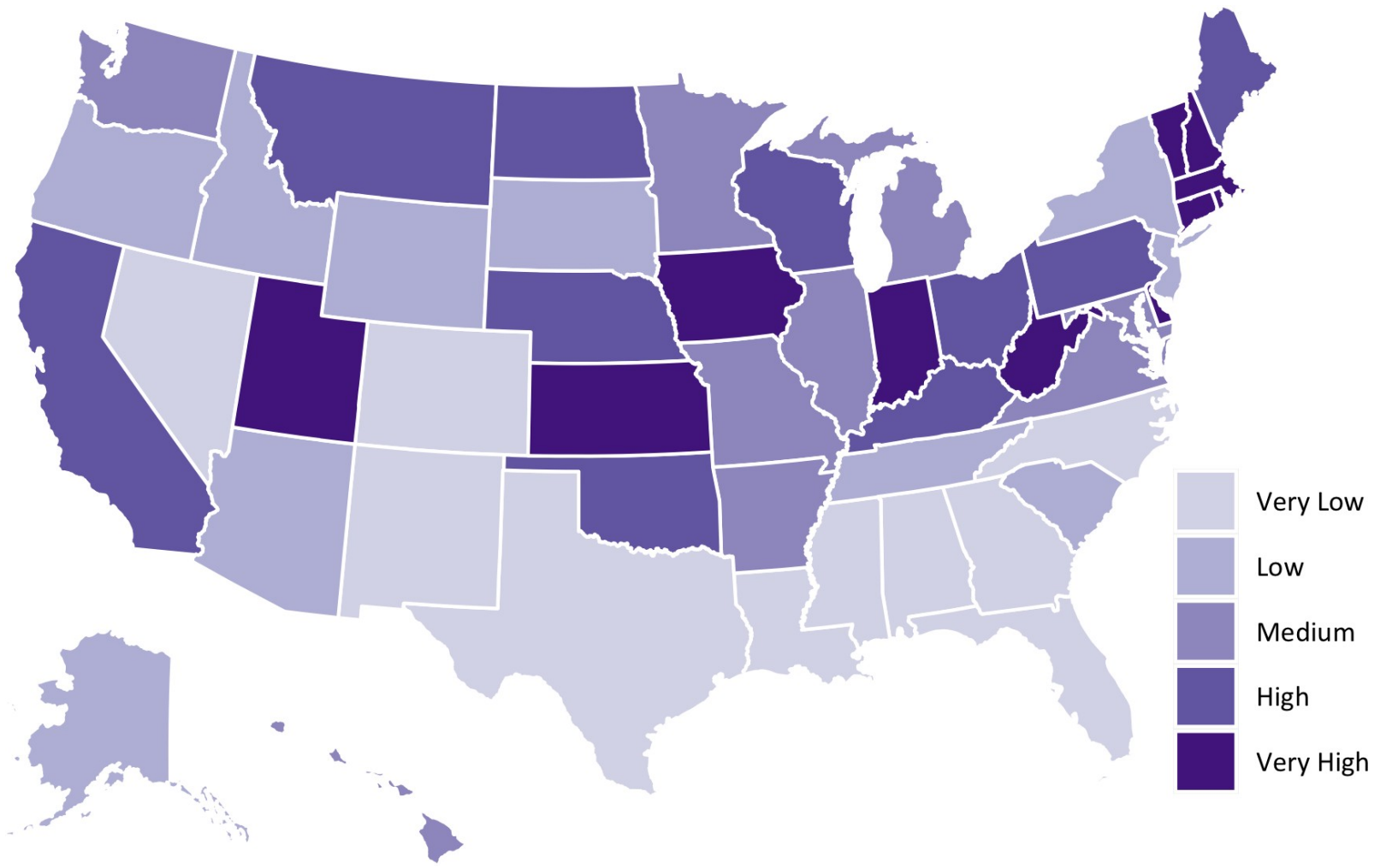

**How is this process statistic calculated?**
Percent of occupied HUs with known status but whose population count was imputed for 2020 minus the corresponding percentage for 2010.

**How is this process statistic interpreted?**
Ideally, the Census Bureau can determine how many people reside in a HU after several visits or with administrative records. But inevitably, these procedures may prove inconclusive. The Bureau may then use information on neighboring HUs, a process known as imputation. This PS reflects quality because imputations may underestimate or overestimate the number of residents, increasing the risk of error.

**Key Points**

- Nationwide, the percent of imputed population counts declined slightly between 2010 and 2020.
- Levels stayed the same or declined in all but a handful of states. In those states, the increases were relatively small. This suggests that one or more census operations precluded the need for count imputation.

**Future Research**

- How are count imputation and coverage error related?
- What occurred in the census process, and especially in field operations, that held the level of count imputation in check, despite the pandemic?

| Rank | State | Percent of occupied HUs with known status but whose population count was imputed for 2020 | Process Statistic 9: Percent of occupied HUs with known status but whose population count was imputed for 2020 minus the corresponding percentage for 2010* |
|---|---|---|---|
| 1 | New Mexico | 0.06 | -0.71 |
| 2 | Washington, D.C. | 0.10 | -0.67 |
| 3 | Georgia | 0.12 | -0.60 |
| 4 | Louisiana | 0.17 | -0.49 |
| 5 | Mississippi | 0.12 | -0.47 |
| 6 | Alabama | 0.10 | -0.44 |
| 7 | Colorado | 0.06 | -0.44 |
| 8 | North Carolina | 0.09 | -0.43 |
| 9 | Nevada | 0.09 | -0.41 |
| 10 | Texas | 0.05 | -0.39 |
| 11 | Florida | 0.07 | -0.38 |
| 12 | South Dakota | 0.03 | -0.34 |
| 13 | Arizona | 0.03 | -0.33 |
| 14 | South Carolina | 0.08 | -0.33 |
| 15 | Tennessee | 0.06 | -0.32 |
| 16 | Wyoming | 0.02 | -0.31 |
| 17 | New Jersey | 0.06 | -0.28 |
| 18 | Idaho | 0.06 | -0.27 |
| 19 | Alaska | 0.05 | -0.25 |
| 20 | Oregon | 0.06 | -0.24 |
| 21 | New York | 0.09 | -0.23 |
| | United States | 0.06 | -0.22 |
| 22 | Washington | 0.04 | -0.20 |
| 23 | Maryland | 0.07 | -0.18 |
| 24 | Arkansas | 0.05 | -0.17 |
| 25 | Illinois | 0.05 | -0.15 |
| 26 | Hawaii | 0.04 | -0.14 |
| 27 | Virginia | 0.04 | -0.13 |
| 28 | Michigan | 0.05 | -0.12 |
| 29 | Missouri | 0.07 | -0.12 |
| 30 | Minnesota | 0.03 | -0.11 |
| 31 | Montana | 0.08 | -0.10 |
| 32 | Wisconsin | 0.05 | -0.09 |
| 33 | Kentucky | 0.08 | -0.08 |
| 34 | Pennsylvania | 0.07 | -0.08 |
| 35 | California | 0.05 | -0.07 |
| 36 | North Dakota | 0.03 | -0.07 |
| 37 | Ohio | 0.05 | -0.06 |
| 38 | Oklahoma | 0.03 | -0.05 |
| 39 | Maine | 0.04 | -0.05 |
| 40 | Nebraska | 0.02 | -0.04 |
| 41 | Utah | 0.05 | -0.03 |
| 42 | Iowa | 0.05 | -0.01 |
| 43 | Indiana | 0.04 | -0.01 |
| 44 | New Hampshire | 0.05 | 0.00 |
| 45 | Connecticut | 0.05 | 0.01 |
| 46 | Rhode Island | 0.10 | 0.01 |
| 47 | Kansas | 0.06 | 0.01 |
| 48 | Delaware | 0.08 | 0.02 |
| 49 | West Virginia | 0.07 | 0.02 |
| 50 | Massachusetts | 0.11 | 0.05 |
| 51 | Vermont | 0.06 | 0.05 |

*States are sorted and colored by the risk of error implied by the process statistic—from lowest risk (rank 1, lightest color) to highest risk (rank 51, darkest color). The bars facilitate comparisons between states, having 0 length for the lowest risk state and filling the entire cell for the highest.

### Process Statistic 10: Group Quarters With Imputed Count

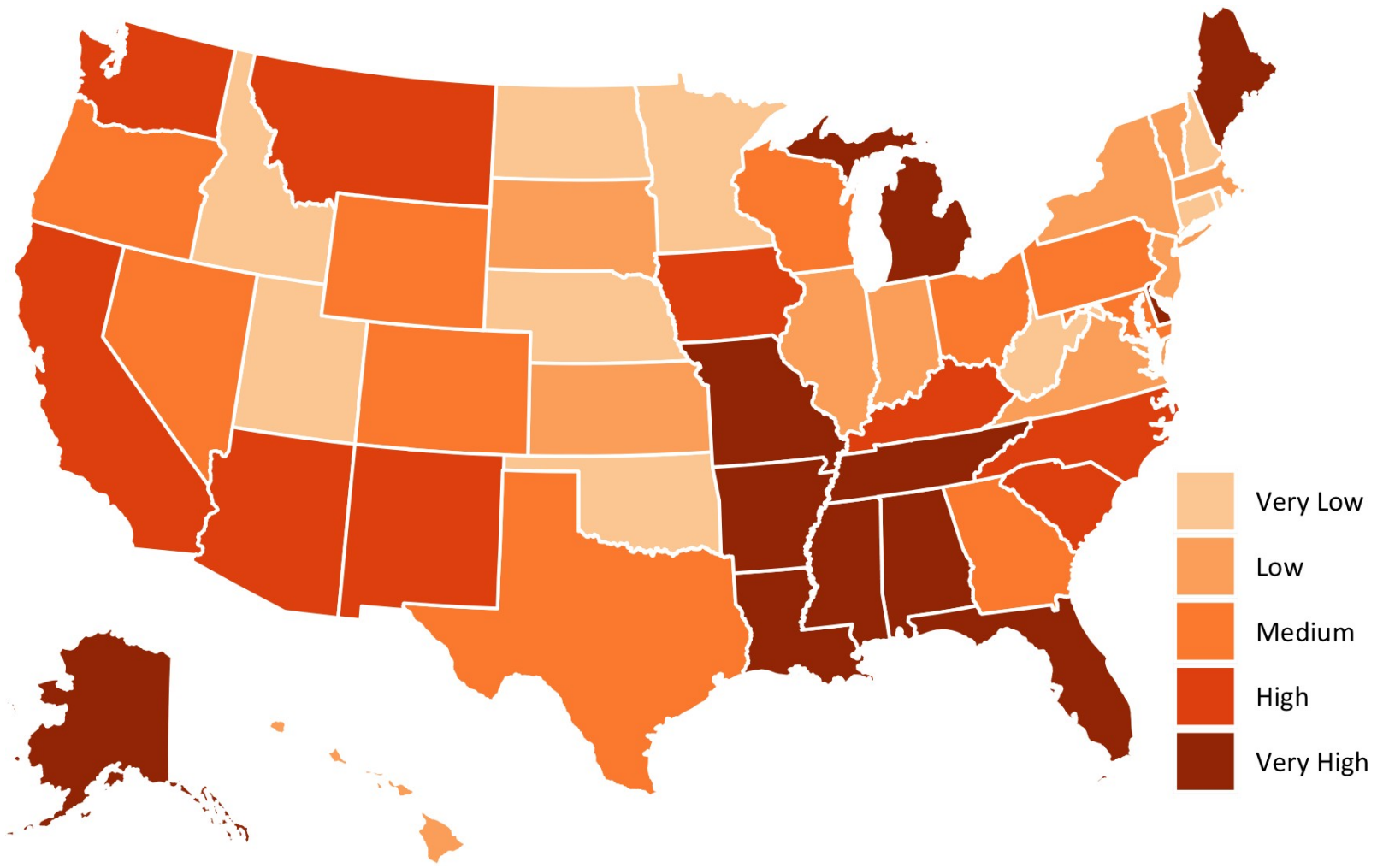

**How is this process statistic calculated?**

Percent of the GQs population that was imputed in 2020.

**How is this process statistic interpreted?**

Ideally, administrators of GQs facilities submit an accurate population count. But the pandemic greatly complicated efforts to count GQs, particularly skilled nursing facilities and college residences. The Census Bureau used statistical methods to predict the population when population counts were not available, a process known as imputation. This PS reflects quality because imputations may underestimate or overestimate the population of GQs, increasing the risk of error.

**Key Points**

- Nationwide, roughly 2% of the Total GQs population was imputed in 2020.
- The percentages varied widely by state, from 0.13% (NH) to more than 11% (DE). DE and MS were unusually high relative to other states.
- The GQs population is small overall, but some states had fairly high levels of count imputation.

**Future Research**

- Did GQ imputation increase coverage error?
- How did the level of imputation vary by GQs type, given the unequal impact of the pandemic?

| Rank | State | Process Statistic 10: Percent of the GQs population that was imputed in 2020* |
|---|---|---|
| 1 | New Hampshire | 0.13 |
| 2 | Rhode Island | 0.21 |
| 3 | Washington, D.C. | 0.26 |
| 4 | Minnesota | 0.32 |
| 5 | North Dakota | 0.32 |
| 6 | Nebraska | 0.41 |
| 7 | Connecticut | 0.50 |
| 8 | Idaho | 0.50 |
| 9 | Oklahoma | 0.53 |
| 10 | West Virginia | 0.53 |
| 11 | Utah | 0.55 |
| 12 | Kansas | 0.57 |
| 13 | Massachusetts | 0.58 |
| 14 | Hawaii | 0.63 |
| 15 | Indiana | 0.64 |
| 16 | New Jersey | 0.68 |
| 17 | Vermont | 0.70 |
| 18 | Virginia | 0.71 |
| 19 | New York | 0.79 |
| 20 | South Dakota | 1.00 |
| 21 | Illinois | 1.04 |
| 22 | Colorado | 1.10 |
| 23 | Ohio | 1.14 |
| 24 | Maryland | 1.23 |
| 25 | Wisconsin | 1.39 |
| 26 | Pennsylvania | 1.60 |
| 27 | Oregon | 1.64 |
| 28 | Nevada | 1.72 |
| 29 | Texas | 1.90 |
| 30 | Wyoming | 1.92 |
| 31 | Georgia | 1.99 |
| 32 | California | 2.00 |
| | United States | 2.01 |
| 33 | Montana | 2.01 |
| 34 | Kentucky | 2.17 |
| 35 | Iowa | 2.25 |
| 36 | Washington | 2.72 |
| 37 | North Carolina | 3.07 |
| 38 | New Mexico | 3.15 |
| 39 | Arizona | 3.19 |
| 40 | South Carolina | 3.49 |
| 41 | Florida | 3.53 |
| 42 | Arkansas | 3.57 |
| 43 | Tennessee | 3.62 |
| 44 | Missouri | 3.95 |
| 45 | Louisiana | 4.34 |
| 46 | Alaska | 4.75 |
| 47 | Michigan | 4.76 |
| 48 | Alabama | 4.81 |
| 49 | Maine | 4.98 |
| 50 | Mississippi | 6.81 |
| 51 | Delaware | 11.26 |

*States are sorted and colored by the risk of error implied by the process statistic—from lowest risk (rank 1, lightest color) to highest risk (rank 51, darkest color). The bars facilitate comparisons between states, having 0 length for the lowest risk state and filling the entire cell for the highest.

### Summary Process Statistic

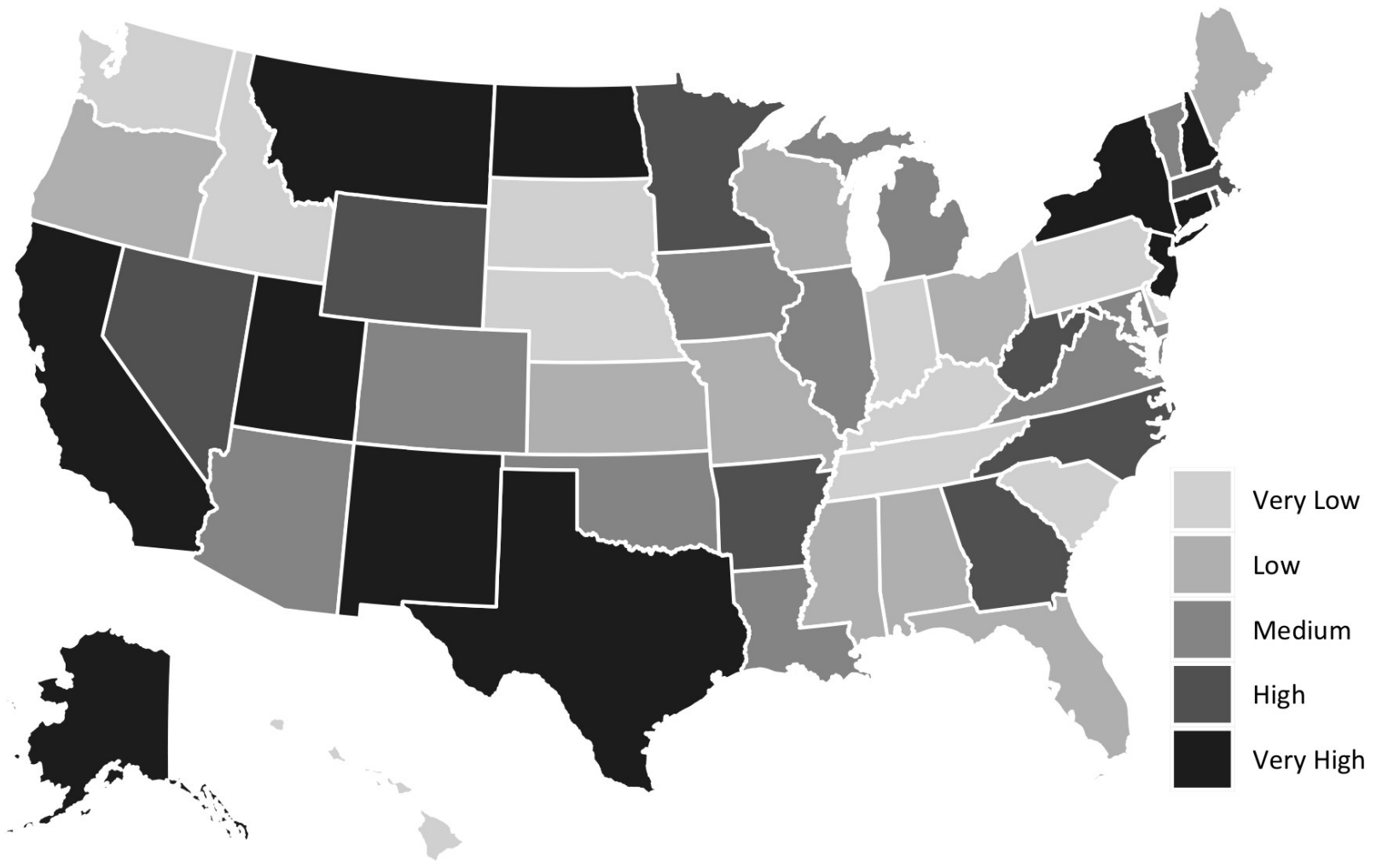

**How is the summary process statistic calculated?** For each PS, states are assigned a number between 1 and 5 (quintile) reflecting risk of error: 1 denotes very low risk, 2 low, 3 medium, 4 high, 5 very high. The SPS is the weighted average of these numbers. Weights are chosen to reflect the proportion of the count at risk.

**How is the summary process statistic interpreted?** The SPS is an overall measure of the error risk from all 10 PSs. A state at higher risk across all statistics will have a higher SPS value. For example, an SPS value of 5 means all 10 PSs are at the highest risk level and a value of 1 means they are all at the lowest risk level.

**Key Points**

- Five highest risk states are AK, NJ, UT, NY, and TX.
- Eight lowest risk states have an average SPS less than 2.

**Future Research**

- Do states with a higher SPS also have higher levels of coverage error?
- What is the best way to summarize the risk that a state's population count will be significantly higher or lower than the true count?
- What are the characteristics of substate areas with higher values for the SPS?

| Rank | State | Summary Process Statistic: Weighted average of the ten process statistics* |
|---|---|---|
| 1 | Nebraska | 1.21 |
| 2 | Indiana | 1.24 |
| 3 | Tennessee | 1.70 |
| 4 | Kentucky | 1.86 |
| 5 | Idaho | 1.93 |
| 6 | Delaware | 1.94 |
| 7 | South Dakota | 1.98 |
| 8 | South Carolina | 1.98 |
| 9 | Washington | 2.01 |
| 10 | Hawaii | 2.11 |
| 11 | Pennsylvania | 2.14 |
| 12 | Wisconsin | 2.27 |
| 13 | Alabama | 2.28 |
| 14 | Mississippi | 2.41 |
| 15 | Ohio | 2.46 |
| 16 | Kansas | 2.47 |
| 17 | Maine | 2.50 |
| 18 | Oregon | 2.58 |
| 19 | Missouri | 2.60 |
| 20 | Washington, D.C. | 2.61 |
| 21 | Florida | 2.63 |
| 22 | Virginia | 2.97 |
| 23 | Arizona | 2.98 |
| 24 | Vermont | 3.01 |
| 25 | Iowa | 3.04 |
| 26 | Michigan | 3.06 |
| 27 | Louisiana | 3.07 |
| 28 | Colorado | 3.09 |
| 29 | Maryland | 3.12 |
| 30 | Illinois | 3.13 |
| 31 | Oklahoma | 3.13 |
| 32 | West Virginia | 3.17 |
| 33 | Wyoming | 3.18 |
| | United States | 3.28 |
| 34 | North Carolina | 3.29 |
| 35 | Arkansas | 3.30 |
| 36 | Nevada | 3.41 |
| 37 | Rhode Island | 3.41 |
| 38 | Massachusetts | 3.50 |
| 39 | Georgia | 3.52 |
| 40 | Minnesota | 3.57 |
| 41 | California | 3.73 |
| 42 | Connecticut | 3.76 |
| 43 | North Dakota | 3.76 |
| 44 | New Hampshire | 3.94 |
| 45 | New Mexico | 4.13 |
| 46 | Montana | 4.14 |
| 47 | Texas | 4.19 |
| 48 | New York | 4.25 |
| 49 | Utah | 4.28 |
| 50 | New Jersey | 4.38 |
| 51 | Alaska | 4.47 |

*States are sorted and colored by the risk of error implied by the process statistic—from lowest risk (rank 1, lightest color) to highest risk (rank 51, darkest color). The bars facilitate comparisons between states, having 0 length for the lowest risk state and filling the entire cell for the highest.

### Decomposition of Summary Process Statistic by Census Phase

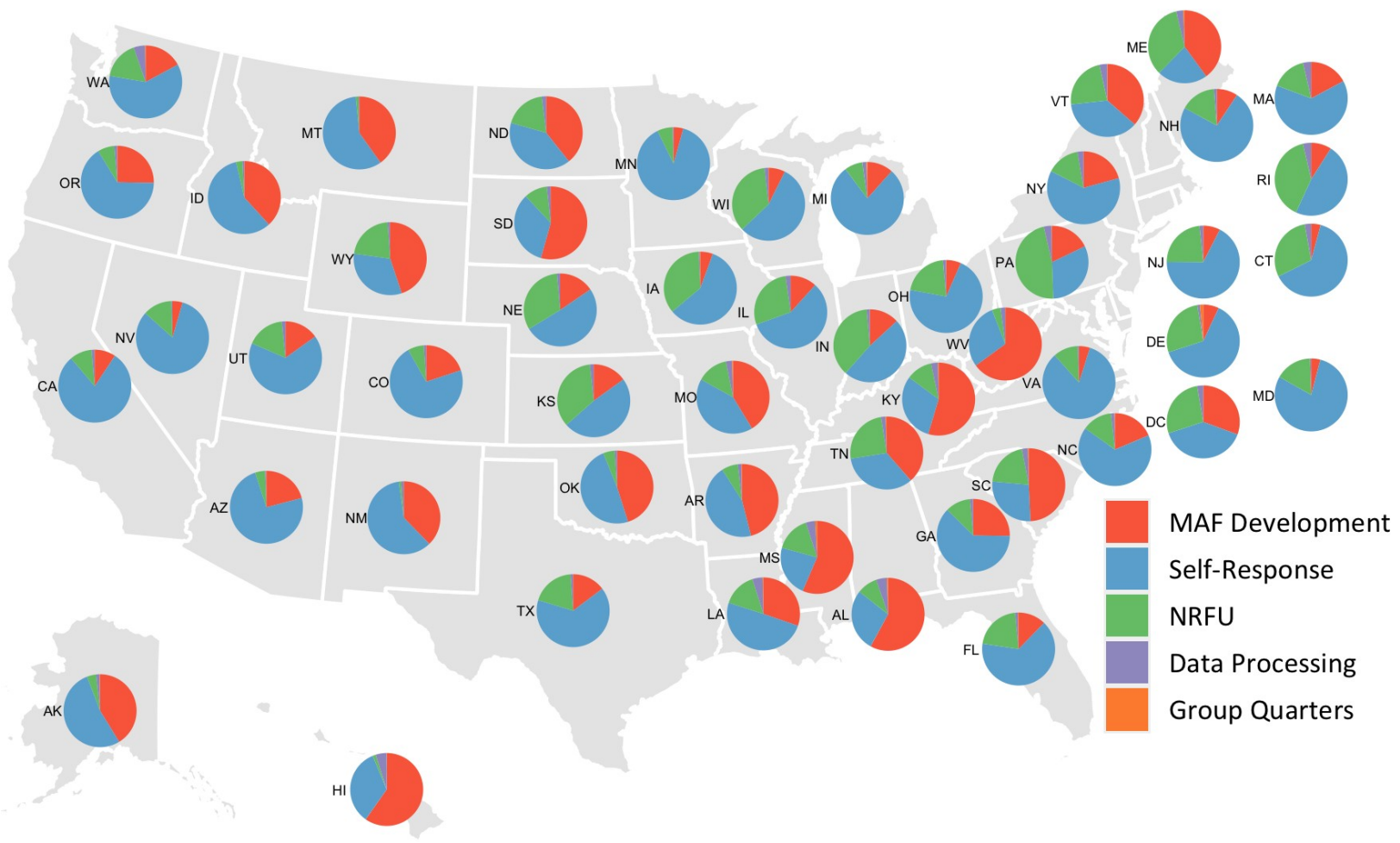

**How is the summary process statistic decomposed?** Rather than summing the risk of error (quintiles) across all 10 PSs as in the previous profile, sum only PSs within the same census phase (i.e., MAF, SR, NRFU). Weight reflects the proportion of the count potentially at risk.

**How is this decomposition interpreted?** The phase-level SPS reflects the contribution of each of the five phases to total census quality. For example, as shown in Table 1, the SR phase includes Non-Matching No IDs, Multiple Responses, and URCs. The SR-level statistic reflects the combined contribution of these three PSs (as a percentage of the SPS) to the overall SPS.

**Key Points**

- The operation producing the greatest risk was Self-Response, accounting for the largest percentage of the SPS for 39 states.
- MAF Development was second greatest, accounting for the largest percentage of the SPS for 12 states.
- Only in PA did NRFU account for the largest percentage.

**Future Research**

- Do the operations that contribute the most to the SPS produce more coverage errors?
- What other approaches might answer questions about how error risks vary geographically?

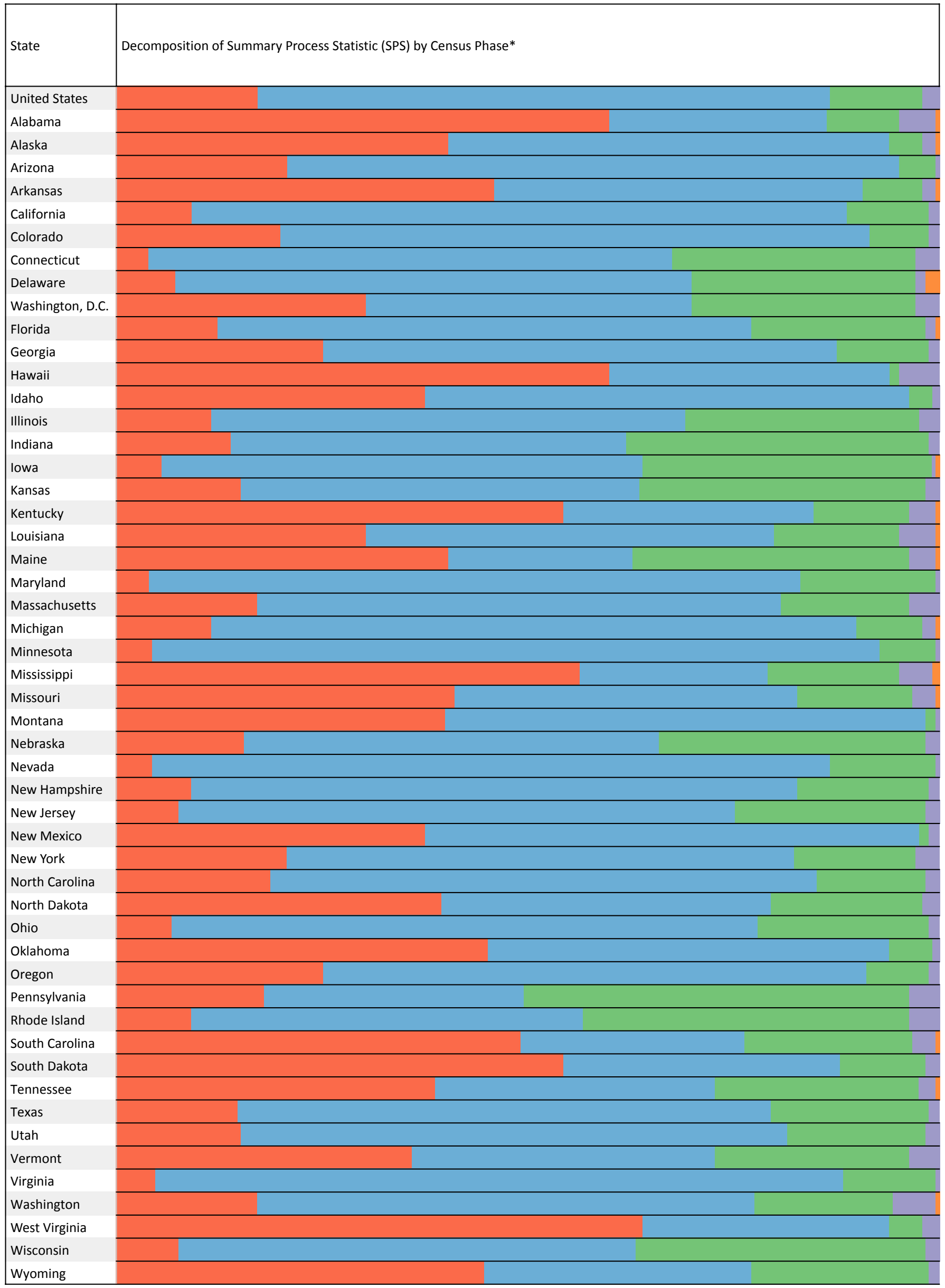

*States are listed alphabetically following the U.S. The width of the second column represents the SPS of each state. Each color represents the relative contribution of the process statistics in each census phase: MAF Development (red), SR (blue), NRFU (green), Data Processing (Purple), and GQ (orange).

# V Methodology

## General Approach

This report attempts to address the question: "what evidence is there that the quality of the 2020 census apportionment counts is less than the quality of the 2010 census counts?" To this end, PSs from five phases of the 2020 census are identified, evaluated, and compared with PSs from the 2010 census whenever possible. Use of the 2010 census as a reference point is not meant to imply that that census is regarded as a standard for perfection. Rather, because the 2010 census was not subject to a pandemic and had fewer other issues than 2020, it can be used as the most recent example of how a "typical" census might perform.

More specifically, this analysis partitions Census Bureau activities into the following five census phases:

1. Master Address File Development (MAF)
2. Self-Response (SR)
3. Nonresponse Follow-up (NRFU)
4. Data Compilation and Processing
5. Group Quarters (GQs)

It then seeks to discover the following:

1. For all phases of the census from MAF development to data compilation and processing, identify activities within each census phase that pose some risk of coverage error at the state level.
2. For each activity, define one or more PSs that reflect potentially elevated error risks in performing the activity.
3. If similar PSs are available from 2010, compare the 2020 and 2010 PSs to determine activities that could have greater error risk (i.e., more opportunities for error in 2020 than in 2010).

For each census phase, an ideal scenario can be envisioned. For example, for MAF Development, the MAF would be complete and accurate; for Self-Response, one and only one resident from each household would complete a census questionnaire; and so on. Deviations from these ideals tend to increase risk of coverage errors.

A total of 10 critical activities are identified with at least one critical activity per census phase. A critical activity is a major census operation designed to mitigate possible errors resulting from deviations from an ideal process. They are associated with error risk in that some deviations from the ideal will not be mitigated successfully and, for those, coverage errors could result. An activity is critical if its successful implementation is critical to census quality. Our analysis assumes that the risk of error increases each time the activity is performed. In other words, errors are more likely to occur as the opportunities for an error are increased.

For each critical activity, one or more PSs are defined at the state level to assess the performance of the activity as it may affect the apportionment counts. All 10 PSs defined in this report are based on the proportions of cases (i.e., HUs, persons or addresses) affected by some critical activity. Six PSs are formed as the difference between the proportions from the 2020 and 2010 censuses for the same critical activity. For example, PS 1 (MAF Addresses Having Imputed Status) is the difference in the proportions of MAF addresses with imputed status between 2020 and 2010. In four cases, the critical activity either did not exist in 2010 or its comparability to 2020 could not be established. Thus, the PS only reflects the performance of the critical activity in 2020.

To the extent that a PS can be associated with the risk of coverage error from an activity, the variation in a PS across states also reflects the variation in error risks from its related activity. For PSs that are differences between 2020 and 2010, positive values imply greater risk from the activity in 2020 than in 2010 while negative values imply lesser risk in 2020. Similarly, for PSs defined for 2020 only, larger values for a state imply large risks from the critical activity in that state compared with other states having smaller values.

All 50 states, Washington, D.C., and the entire United States are ranked according to each PS. The higher the value of the PS, the higher the ranking. In addition, an SPS is created that may be regarded as an indicator of average error risk from all 10 critical activities. The SPS is formed by first replacing each PS by its quintile ranking and then taking the weighted average of the 10 quintile ranked PSs. Here the weight applied to a PS for a state is proportional to the number of persons in the state count affected by the critical activity.

*Error Risks and Process Statistics*

It is important to understand that PSs are not error rates and should not be interpreted as indicators of error. The process statistic is a well-known concept in the survey quality literature. It is defined as any quantity computed from operational data that may be related to the performance of the operation (see, for example, Biemer & Lyberg, 2003, Table 7.2). A large value for a PS does not mean that errors occurred in an operation. Rather, the PS is intended to reflect the error risk its underlying activity presents to data quality. A higher PS value simply implies more opportunities for error and, thus, a greater chance that some of those opportunities resulted in an error.

It is also important to understand that errors can be offsetting when aggregated to the state level (i.e., undercounts may offset overcounts to some unknown extent when the apportionment counts are tallied). The PSs in this report are not intended to provide any information on the risk of net coverage error, which is more relevant than gross error for the purposes of assessing apportionment count accuracy. This is another important reason that interpreting error risks at the state level as evidence of actual error in state counts is inappropriate.

To illustrate the proper interpretation of a PS, consider the process statistic defined for the use of proxy respondents – a critical activity that is performed on the occupied HUs in the NRFU universe. When a household member cannot be contacted or refuses to provide information on an HU, a neighbor or other knowledgeable informant, referred to as a "proxy" respondent, may be consulted to supply the required information. The PS for proxy respondents is the difference between 2020 and 2010 in the proportion of occupied HUs where a proxy respondent was used to determine the count of persons living at an address and possibly other characteristics about the residents. The Census Bureau's analysis of the 2010 Post-Enumeration Survey results suggests that proxy interviewing poses some appreciable risk of coverage error because proxies may not be sufficiently knowledgeable about the household to provide accurate information. Therefore, the proportion of proxy interviews is an informative PS for assessing the error risk associated with NRFU proxy responses. The PS provides a count of how often a proxy was used during NRFU as a proportion of occupied HUs. Larger proportions suggest greater potential for proxy error than smaller proportions.

Activities like proxy interviewing, status imputation, count imputation, and GQ imputation, carry a relatively high risk, supported by the census literature, that a household count could be in error. For other activities like the use of administrative records, resolving multiple responses, and MAF revisions, the evidence of error risk is weaker and more speculative. Nevertheless, these activities involve decisions that may be difficult in many situations and thus subject to error. If the proportion of cases affected by these activities is larger in 2020 than in 2010, inferring greater error risk in 2020 from these activities than in 2010 is still justified. This is because the interpretation of the PS does not depend on whether the actual error rates (i.e., the percent of opportunities for error that actually result in error) are large or small. What matters is whether the activity error rates are about the same for 2010 and 2020. If this is true, then it follows that an increase in the PS suggests an increase in error risk.

Finally, note that this evaluation does not represent the definitive statement on the quality of the 2020 census. Rather, the Census Bureau is expected to release the results of its Post-Enumeration Survey as it always has in recent censuses. This survey is conducted as a follow-up to the census and will provide estimates of census coverage error (i.e., overcounts and undercounts) for all states, many substate areas, and various subpopulations. The first Post-Enumeration Survey results are expected in the first quarter of 2022. Nevertheless, the PSs in this report provide a glimpse at the operations in the census that we hope will be informative and useful for understanding the quality associated with the state population counts since they may affect apportionment.

*Quintile Transformation of PSs*

Although raw values for a single PS may be useful for a state-by-state comparison of the risks for a particular activity or phase, they are not useful for depicting patterns across different PSs within a state because the measurement scales may be very different. For example, an important question in this evaluation is whether a particular state's PSs are predominantly high (indicating high overall risk) or low (indicating low overall risk). Because each PS may have a different base (e.g., addresses, HUs, or persons) and a different range of values, comparing their raw values can be both confusing and misleading. To address this issue, we decided to standardize the PSs so that their relative magnitudes are more meaningful and comparable across activities. Thus, for our primary analysis, we replaced each PS's raw value by its quintile (i.e., 1, 2, 3, 4, or 5) relative to the other 52 entities (i.e., 50 states, Washington, D.C., and the United States), where each quintile represents 20 percentage points. Quintiles for a PS can be computed by first ranking the PS from smallest to largest. The smallest 11 states are assigned the value 1, the next 10 the value 2, the next 10 the value 3, the next 10 the value 4, and the final 11 the value 5. As an example, the value of PS for Multiple Responses is 15.18 for Alabama, which is assigned the value 1 because it is among the smallest 11 states for this PS. In this same manner, all the PSs are converted to the numbers 1, 2, 3, 4, or 5 corresponding to their respective quintiles. Furthermore, in each case, the higher a PS's quintile ranking, the higher the error risk is for the census activity measured by that PS relative to the 50 other states. The original values of the PSs have been documented in this report and can also be found in Appendix A (see Table A1.)

*Type of Enumeration Areas*

Unless otherwise stated in our analysis, the PSs are defined across all census type of enumeration areas (TEAs) for both 2020 and 2010. In 2020, four TEAs were defined. About 95% of households received their census invitation in the mail (TEA 1) and almost 5% received their invitation when a census taker dropped it off at their home in so-called update leave (UL) areas (TEA 6). The remaining less than 1% of areas—mostly remote—were counted in person by a census taker instead of being invited to respond on their own (TEAs 2 and 4). Although TEA 6 accounts for only a small percentage of all HUs in the nation, in some states a much higher percentage of housing is in these areas (see 2020 Census: Type of Enumeration Area (TEA) Viewer). It is also important to note that states having a higher percentage of UL or Update Enumerate (UE) areas may have different error risks than states where only a small fraction of the enumerations used these data collection methods.

**Process Statistics by Census Phase**

This section describes 10 critical activities examined in this review along with their respective PSs as listed in Process Statistics at a Glance. The section is organized by the five phases of the census identified in the previous section. After discussing each critical activity and its corresponding PS, we describe the SPS that combines all 10 PSs into a single PS.

*MAF Development Phase*

1. MAF revisions (2020 only)

Process Statistic: Percent of all MAF addresses that were either added or deleted during the 2020 census data collection period.

The foundation for the decennial census is the list of addresses used to mail packets with a request to respond to the decennial census. To be included in the census, every person must be linked to a physical address of an HU or GQs facility. This list comes from the Census Bureau's MAF, a repository for all information collected over time for an address—a kind of longitudinal history of addresses associated with an HU. The Census Bureau does biannual updates of the MAF using data from the U.S. Postal Service's Delivery Sequence Files, which is used for mail delivery. In addition, the Census Bureau solicits address updates from tribal, state, and local governments as part of several geographic partnership programs throughout the decade. The MAF is a dynamic list that reflects the ever-changing inventory of housing in the United States.

The MAF errors that are the biggest concern for the apportionment counts occur when addresses on the MAF that are not living quarters are classified that way or addresses that are associated with living quarters are missing from the MAF. Census Bureau procedures during the enumeration phase of 2020 census operations dictate that addresses on the MAF that are not living quarters should be deleted from the MAF (referred to as "deletes") and living quarters that are not on the MAF should be added to the MAF (referred to as "adds"). However, as in any extremely large and complex operation, errors can occur. The most common errors are the following:

1. Non-living quarters listed on the MAF are classified as occupied HUs.
2. Non-living quarters not listed on the MAF are added as living quarters during data collection.
3. Living quarters on the MAF are classified as non-living quarters.
4. Living quarters missing from the MAF are not discovered during data collection.

Errors 1 and 2 can lead to overcoverage, while errors 3 and 4 can lead to undercoverage.

An ideal PS for the MAF is one that captures all four error risks. The Post-Enumeration Survey, as previously noted, will provide estimates of the rates at which these errors occur, but until then, only indirect risk measures are available.

PSs that reflect the risk of MAF errors is a relatively new concept. Indeed, metrics that reflect assessing the risk of errors 1 through 4 have not been reported by the Bureau before, to our knowledge. The literature on frame coverage error provides some guidance (see, for example, Biemer & Lyberg, 2003 [Chapter 9]); however, those quality evaluation methods usually require a benchmark file that is more accurate than the frame under consideration. Only then can errors 1 through 4 above be evaluated with sufficient accuracy. Unfortunately, no list of addresses for the United States is currently available that is more complete or of higher quality than the MAF that could serve as a benchmark for measuring MAF coverage errors.

For the purposes of this review, we developed the concept of MAF Revisions. Revisions refers to how much the MAF changed during census data collection (i.e., the number of units deleted and added). Figure 1 contains a Venn diagram where the circle to the left represents the MAF at the beginning of data collection (on Census Day April 1, 2020) and the circle to the right represents the MAF at the end of data collection (October 15, 2020), referred to as MAF1 and MAF2, respectively. The intersection of the circles represents addresses that we classified as living quarters on both MAF1 and MAF2. The part labeled MAF1 only are addresses reclassified "not living quarters" on MAF2 (i.e., deletes) and the part labeled MAF2 only are addresses classified as "living quarters" on MAF2 that were not classified as such on MAF1 (i.e., adds). Thus, the MAF revision of addresses PS (referred to as MAF Revisions) is the proportion of all addresses in the combined MAF1 and MAF2 files that are not listed on both MAF1 and MAF2 (i.e., the proportion of addresses in MAF1 only plus MAF2 only).

MAF Revisions measures the change to the MAF at two timepoints in 2020: April 1 to reflect "Census Day" and October 15 when census field work ended. The changes in the MAF between these two timepoints reflect addresses that were deleted and added to the MAF. The greater the percentage of MAF revisions, the greater the number of opportunities for error (i.e., error risk). Thus, MAF Revisions reflects the number of add/delete decisions made and thus, the number of chances that error types 1 through 4 could occur.

*Figure 1. A Venn Diagram Representing the MAF on April 1 (MAF1) and October 15 (MAF2)*

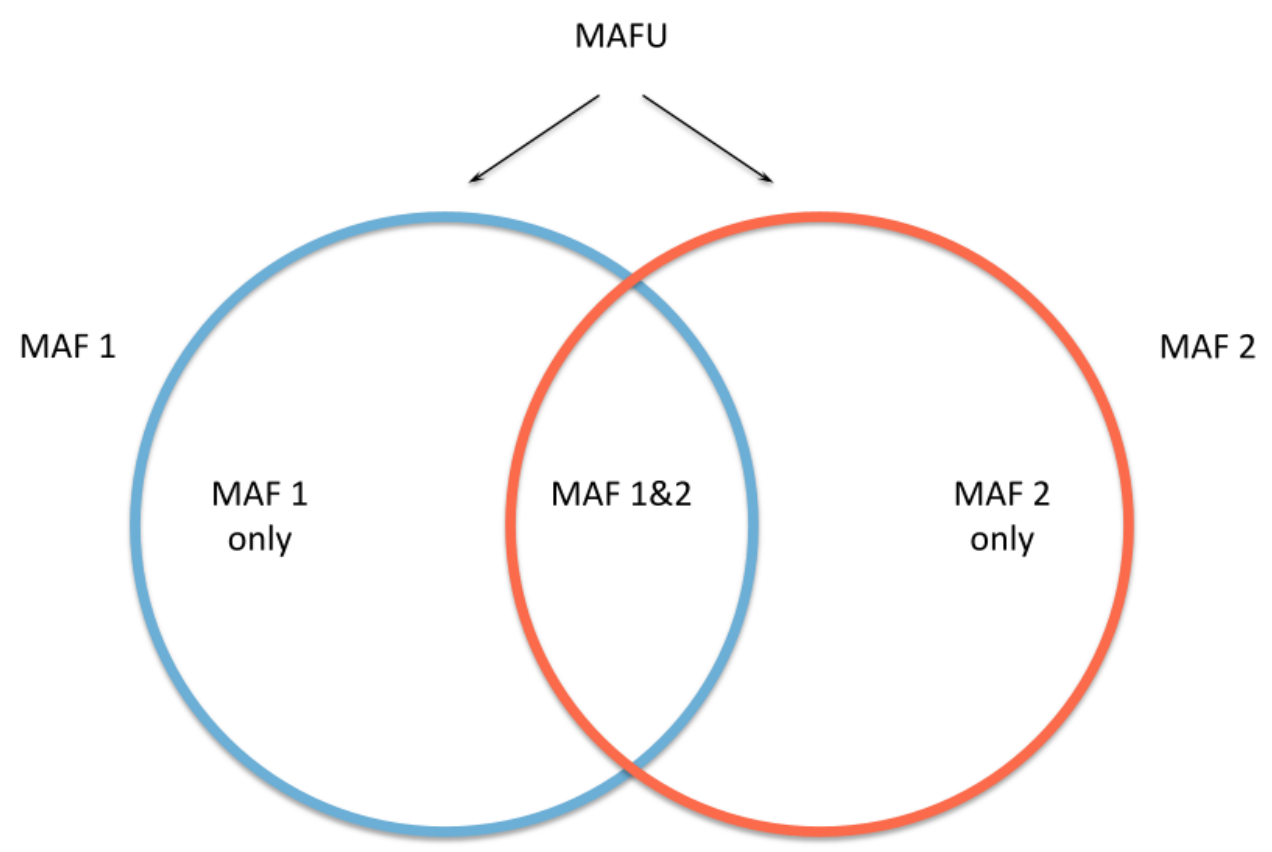

Consideration was given to the development of an equivalent MAF revisions PS for 2010; however, this was not possible because the MAF development processes for 2020 and 2010 were so different. For example, rather than the full address canvassing operation implemented in 2010, about 65% of the MAF used an "in-office" canvassing and review approach. Field address canvassing was implemented only in difficult-to-canvass areas that required ground truth data. In addition, the initial MAF in 2020 (MAF1 in Figure 1) contained many addresses whose validity was equivocal because of conflicting information from various administrative data sources including the U.S. Postal Service. Thus, many addresses on the 2020 MAF likely would not have been included on the MAF in 2010. These differences between 2020 and 2010 made the creation of a comparable PS for 2010 infeasible.

*Self-Response Phase*

2. Matching census returns with no IDs to the MAF (2020 only)

Process Statistic: The percent of 2020 census HUs submitting questionnaires without census IDs and no matching address was found on the MAF.

A census ID is a unique identifier (somewhat like a Social Security number) that is assigned to every address on the MAF. Every mailed letter requesting self-response included this ID. However, to increase the self-response rate, the Bureau allowed respondents to submit a questionnaire with an address but without a census ID. In that case, the Census Bureau used the respondent's address and other identifiers to match the responding household to a MAF address. The non-ID procedure can work well for returns that have a standard city-style address (i.e., number, street, apartment) and can be readily linked to the Census Bureau's MAF. But it may not work as well for returns that have nonstandard addresses (e.g., when apartment numbers do not formally exist in some cities and towns) that cannot be readily identified on the MAF. The same is true for rural areas where addresses may not have standard or easily identified labeling. Many governments use the Geographic Support Program or Local Update of Census Address (LUCA) program run by the Census Bureau to update the MAF prior to the census to include these types of addresses with appropriate labels.

Although allowing non-ID submissions made it easier for people to respond to the census, it also resulted in returns whose addresses did not match any address on the MAF. The non-ID nonmatching returns resulted in extra work both in the office and in the field to resolve these cases. This problem is unique to the 2020 census because active promotion of non-ID returns was not a feature of the 2010 census. Thus, the presence of non-ID returns, particularly those with nonmatching addresses, increased the opportunities for error in 2020 compared to 2010.

3. Resolving multiple responses (2020 vs. 2010)

Process Statistic: Percent of HUs having two or more questionnaires for 2020 minus the corresponding percentage for 2010.

The major single cause of multiple returns is duplicate questionnaires (i.e., households with the same identification number or address responding multiple times with possibly different responses). The 2020 census provided multiple options for responding: mail, telephone, Internet (for the first time), or enumerator-assisted interviews. In addition, the same household could submit returns with or without an ID. Returns submitted with the same ID could be readily identified during data processing, where a process referred to as the Primary Selection Algorithm decided which return to count.

Identifying multiple or duplicate returns where one or more returns were submitted without an ID required more effort, particularly if the same household used slight variations of the same address. In many of these situations, additional fieldwork was required by the Bureau to verify these addresses and reconcile conflicting address information, often under tight time constraints. The time-constrained field period may have increased the risk that enumerators would falsify information in the field to save time. It may have also required the Census Bureau to perform additional investigations and evaluations to reconcile inconsistent data during the post-data collection processing phase. Duplicate submissions for the same household that are not detected could increase population counts in some states more than others, leading to differential coverage error by state that could skew the apportionment results.

However, a majority of multiple responses resulted for reasons other than respondent-generated duplication. These include:

- Duplicate questionnaires created when an original return's accuracy was deemed to be unacceptable (e.g., questionnaires suspected of being falsified).
- HUs that were ultimately classified as GQs.
- Invalidated records, for example, continuation forms that were later linked to their parent form.
- Multiple returns and dummy records generated by the Bureau's systems to address various data processing issues.
- Partial internet responses that were deemed ineligible when a more complete response was received.

Regardless of the causes of a multiple return, its handling and resolution may still be viewed as an error risk. In addition, the above reasons for multiple returns also existed in 2010. Thus, an increase in this activity in 2020 may not only suggest greater respondent-generated duplication, but also greater falsification of questionnaires, system problems, data processing issues, and so on. Although each individual cause may not carry the same risk to count accuracy, they all pose some appreciable risk. In addition, a large increase in the volume of multiple returns in 2020 should raise concerns that error risks associated with multiple returns may have also increased.

4. Reassignment of college students with usual residence elsewhere (2020 vs. 2010)

Process Statistic: Percent of occupied HUs with two or more people, where one or more occupant indicated their usual residence was at college, minus the corresponding percentage for 2010.

Population relocations were particularly more frequent for college students since many campuses closed their dormitories because of the pandemic. Along with those living off-campus, many students decided to relocate to the homes of parents, relatives, and friends. These so-called URCs persons created additional challenges for the Census Bureau, especially in locations with high infection rates. Thus, the risk that college students could be counted in multiple locations was greater in 2020 than in 2010. The URC PS captures that risk.

The URC PS is the number of HUs with one or more URCs divided by the number of households of size 2. The justification for this divisor is that, although it was asked for internet and phone respondents, the URE question was not asked for single-person HUs in the paper questionnaire. In addition, households within which the sole resident is a displaced college student are likely to be quite rare compared to their prevalence in households of size 2 or more. Thus, using the total number of HUs as the divisor would artificially attenuate the impact of URC on the enumeration process.

*Nonresponse Follow-up Phase*

5. Proxy response (2020 vs. 2010)

Process Statistic: Percent of 2020 census occupied HUs units whose census count was obtained from a proxy respondent during the NRFU phase of the census minus the corresponding number from the 2010 census.

Data from the 2010 Census Coverage Measurement program have shown that well over 95% of households that self-responded were correctly enumerated. The comparable estimate for households enumerated through an interview with a household member was correct at 93%. This compares with just 70% for proxy respondents and 68% correct enumerations for situations where the type of respondent was unknown. These data suggest that the percentage of returns collected by proxy reflects an important source of error risk. If the proxy rate is greater in 2020 than in 2010 for some state, then the risk of coverage error from this activity is greater in 2020 than in 2010.

6. Count only HUs (2020 vs. 2010)

Process Statistic: Percent of HUs where only a population count was obtained for 2020, minus the corresponding percentage for 2010.

The percent of households where the only information available was a population count could be a sign of resistance among household informants who are only willing to provide the minimal amount of information. It could also suggest a greater willingness of enumerators to accept the minimum rather than pursuing more complete information. This could be the case when enumerators are under pressure to close out their assignments. In either case, it can be regarded as an informative PS for assessing error risk. For example, it is difficult to know if households that provided only a population count are aware of and followed the somewhat complex rules for determining how many persons resided at the address on Census Day. URCs and other occupants whose usual address is elsewhere may have been counted. Other persons who are away but still use the address as their usual residence may not have been counted.

7. Use of administrative records (2020 only)

Process Statistic: Percent of occupied HUs enumerated by administrative records for 2020.

Administrative records played a much smaller role in the 2010 census compared to 2020, although some matching of census responses to administrative records was done in 2010 as a way of resolving potential undercounts of persons reported on the census form as part of Coverage Follow-Up checks. For this reason, the PS for the administrative records use activity only reflects their use in the 2020 census, not the difference between the 2020 and 2010 censuses. Of the more than 2.3 million persons matched to administrative records, counts were determined in the 2010 Census Coverage Measurement program to be correct more than 95% of the time. However, this sample was limited and census tests completed to prepare for the 2020 census revealed flaws in the ability of administrative records to determine occupancy status. In fact, these tests led to a one-visit requirement to confirm occupancy status in 2020. Thus, any definitive conclusions about the accuracy of administrative records must await the results of the Post-Enumeration Survey due out in early 2022. Nevertheless, because administrative records have been shown to be generally less accurate than self-response for determining household size, the use of administrative records is a critical activity in our analysis.

*Data Compilation and Processing Phase*

8. Status Imputations (2020 vs. 2010)

Process Statistic: Percent of MAF units whose status was imputed in 2020, minus the corresponding percentage for 2010.

Sometimes no information is available for an address, even after multiple visits. It is critically important to know whether the address is an occupied or vacant HU, a commercial building, or nonexistent. When the occupancy status or very existence of HUs cannot be determined, statistical imputation is applied, using the attributes of HUs in the neighborhood or other surrounding areas to assign a status to the address in question. The risk of misclassifying an address using imputation is high and can lead to both undercoverage and overcoverage.

9. Count Imputations (2020 vs. 2010)

Process Statistic: Percent of occupied HUs with known status, but whose population count was imputed for 2020, minus the corresponding percentage for 2010.

As a last resort when a building or apartment is determined to be an occupied HU, but there is no reliable source of information about its number of residents, its population count is imputed. Imputation is a process whereby the missing count at an address is obtained from other similarly occupied HUs within the same general vicinity and whose count is known. Although the population count for the unit in question and its “donor” may be quite similar, there is still an obvious risk that the imputed count may be in error.

*Group Quarters Phase*

10. Group Quarters Count Imputation (2020 vs. 2010)

Process Statistic: Percent of the GQs population that was imputed in 2020.

In addition to the designation of addresses as GQs on the MAF, the GQs operation for the 2020 census is conducted using advance visits. These visits were done in January and February 2020, before the onset of the pandemic. Along with input from the LUCA program, the Bureau began the census with a good initial list of facilities and well-developed plans for capturing GQs data. Once pandemic restrictions took hold, however, the enumeration of GQs in a number of categories was disrupted, most notably in college dorms, skilled nursing facilities, and the service-based enumeration (SBE), which was initially scheduled for the end of March 2020.

Efforts to enumerate the population in various GQs categories took place over the course of the year, including the SBE September 22 through 24 and much work in the post-data collection phase of the census. Moreover, when the Census Bureau was still on a December 31 timeline for apportionment numbers to be delivered, a decision was made not to do a Count Review of GQs by members of the Federal-State Cooperative on Population Estimates—a check using local lists of facilities. Once the December 31 deadline was moved, the Census Bureau mounted an effort to “fill in the holes” and avert a potential undercount of GQs in post-data collection processing. Calls to GQs facilities and extended use of administrative data from a variety of sources were deployed in the late stages of data processing. For the first time, imputation was used as a method to determine counts of persons in GQs. Thus, the level of count imputation used to create the GQs population is included as a measure of risk. Because GQ imputation was not used in 2010, this PS only reflects the 2020 census activity.

*Summary Process Statistic*

As previously noted, a major goal of this work is to distinguish states by their risks for errors in the apportionment counts, especially as they exceed the corresponding risks in 2010. This goal is facilitated by using the SPS. The SPS combines all the PSs of interest into a single measure for each state after the PSs have been converted to quintile ranks. The SPS is intended as a measure of the total error risk to the accuracy of a state’s count across 10 critical activities of interest. Note that a measure of total error risk could be formed by simply averaging the PSs of interest. However, some activities, such as address status imputation, affect a relatively small proportion of a state’s count, while other activities, such resolving multiple responses affect a much larger proportion of the count. To account for this variation in effects across critical activities, the PSs are weighted according to their potential impacts and then summed to produce the SPS.

The SPS can be compared across states to identify states having greater and lower error risks. However, another useful way to view the SPS is to decompose it by the five census phases—MAF Development, Self-Response, NRFU, Data Compilation and Processing, and Group Quarters. The contribution of each component to the SPS can be expressed as a percentage of the SPS. For example, suppose the SPS is 4.1 for some state. Suppose further that weighted PSs within each of the five phases are 0.5, 2.8, 0.6, 0.1, 0.1, respectively. Then the contribution to SPS for MAF is 0.5/4.1×100% = 12.2%, for Self-Response is 2.8/4.1×100 = 68.3, and so on. Thus, the SPS decomposition by census phase provides a means for assessing the impact of each phase on total census quality.

*Weighting the Process Statistics*

The purpose of PS weighting is to produce a summary measure of the total error risk for the 2020 census by ensuring that activities affecting more cases carry more weight than activities that affect fewer cases. A PS’s weight is derived by estimating the proportion of the state count that is affected by the census activity underlying the PS, for example, the percentage of occupied HUs that submitted two or more questionnaires as a percent of all occupied HUs. Separate weights are computed for each PS for each state (i.e., 510 weights in all). So that the resulting SPS is also a number between 1 and 5, the weights are scaled so that their sum is 1. These scaled weights are applied to their respective PSs before summing. In other words, the SPS is equal to the PS scaled weight times the PS summed over all 10 PSs.

For example, as shown in *Table A2* in the Appendix, the weight for Status Imputations for Alabama (AL) is 1.09. The sum of all 10 weights for AL is 54.32. Therefore, the scaled weight for AL is 1.09/54.32 or 0.02. Repeating this for all 10 PSs for AL will result in a scaled weight sum of 1. In addition, the quintile ranking for Status Imputations for AL is 5 and thus its weighted contribution is 0.02×5 or 0.1. Repeating this for the remaining nine PSs for AL and summing them results in a SPS of 2.38.

One additional adjustment is applied to the weights before they are used to compute the SPS for PSs that are differences between 2020 and 2010. It happens that, for some states, the difference between 2020 and 2010 could be negative indicating that 2010 exceeded 2020 in terms of the error risk reflected in the PS. When this occurs, the weight is set to 0 for the state indicating that additional impact on the 2020 count relative to 2010 is 0. For example, the weight for the proxy response PS is the percent of occupied HUs whose census count was obtained by a proxy (scaled as previously described). The proxy response PS is the difference in this percent between 2020 and 2010. If the difference is negative for a state (indicating greater proxy use in that state for 2010 than in 2020), the weight is set to 0 so that the SPS shows no increased risk because of proxy respondents for that state.

*Figure 2* shows the range and mean of the scaled weights for each of the 10 PSs.

*Figure 2. Mean and Range of the SPS Scaled Weights for the 10 Process Statistics*

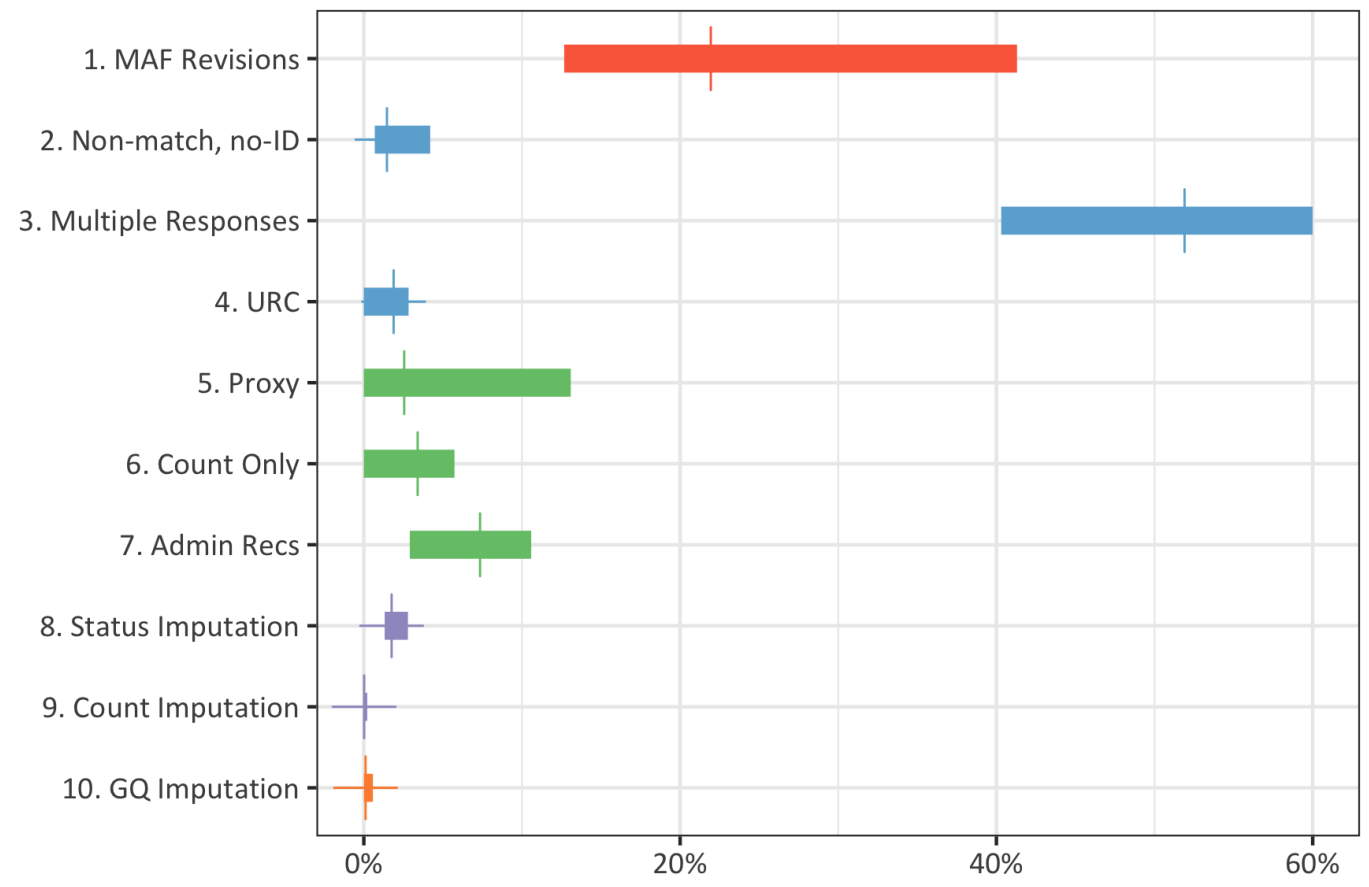

Three points can be made about this figure.

1. Two PSs dominate the SPS: Multiple Responses and MAF Revisions

2. Three PSs have very little effect on SPS: Status Imputations, Count Imputations, and GQ Imputations

3. There is considerable variation across states for several PSs, suggesting that the impact of activities on total risk are highly dependent upon location.

*Interpreting the SPS*

The SPS essentially measures the difference in error risks of the 2020 census compared to the 2010 census. For example, one could interpret an SPS of 3.0 for a state, say State A, as meaning that the error risk for State A is greater in 2020 than in 2010 and, further, the error risk difference is in the middle range among states. That is, about half the states have a greater difference than State A and about half have a lesser difference.

Now suppose State B has an SPS of 4.1. This suggests that (1) State B has greater error risk in 2020 than it did in 2010; (2) the error risk difference for State B is in the top 20% of all states, and (3) its error risk difference exceeds that of State A.

Recall that the weight for a PS is 0 whenever the difference between 2020 and 2010 is negative, which implies error risks were greater in 2010 for that PS's activity than in 2020. Thus, it is possible that SPS could be very small (say close to 1). How would one then interpret the following situation? Suppose that (1) the range of SPS across states is quite small and the largest value of SPS is also quite small. For example, suppose the SPS range is 1.5 to 2.0 for all 51 states and Washington, D.C. Then, one can conclude that the 2020 census differed from 2010 by approximately the same amount for all states, but also, the difference was quite small. In other words, the quality risks for the 2020 census and the 2010 census are about the same.

Finally, it is possible that the difference in error risks for a state in quintile 1 and a state in quintile 5 is very close, even though the state in quintile 1 is in the bottom 20% and the state in quintile 5 is in the top 20% of states. For example, the PS Status Imputations has a range of 0.3 to 1.4. This PS has the smallest range among the 10 PSs. This means that the lowest state differs from 2010 by 0.3 points and the highest state by 1.4 points regarding status imputations. Although this may not seem like a large range, it is still true that the highest state has about 4.7 points more status imputations than the lowest state. In terms of error risk, this may still be an important finding for 2020.

It should be noted that the status imputation PS has a relatively small weight, which limits its influence on the SPS. In fact, for the 10 PSs in our analysis, larger weights are accompanied by wider ranges and smaller weights by smaller ranges of PS values. Thus, the range in values of SPS across states tends to be driven by the PSs with the larger weight.

## VI Summary and Conclusions

Given the difficult circumstances surrounding the 2020 census, data users are concerned about data quality and about potential threats to the counts that affect their share of political representation and access to resources over an entire decade. These concerns were reflected in the publication of the 2020 Census Quality Indicators Task Force report: 2020 Census Quality Indicators: A Report from the American Statistical Association (see ASA Board Releases 2020 Census Quality Indicators), when it recommended that qualified external researchers should be granted access to the data by the Census Bureau to help conduct analyses to assess the quality of the 2020 census.

The analysis in this report is in response to that recommendation. Specifically, our analysis examines 10 PSs associated with activities that are critical to 2020 census quality. The analysis provides important insights regarding the quality of the 2020 census for the state counts. However, the analysis does not provide a definitive statement about 2020 census quality. That will hopefully come in the first quarter of 2022 when the Census Bureau begins releasing results from the Post-Enumeration Survey that will provide estimates of coverage error rates. Still, the current report provides a glimpse at the 2020 census operations. PSs have much to say about the origins of possible error. Thus, they are a necessary but not sufficient step toward a more comprehensive assessment of 2020 census quality.

The next section answers nine questions that we believe are fundamental to understanding 2020 census quality and how it differs from 2010 census quality. Each question is posed followed by an answer that is based on the data in this report.

**Answers to Key Questions Regarding the Quality of the 2020 Census**

**1. What can be said about the quality of the 2020 census compared to the 2010 census?**

The pandemic adversely affected many aspects of life in the United States and the census was no exception. Hurricanes in the south and wildfires in the west further changed the lives of many Americans. These events created a very challenging environment for conducting the 2020 census—much more so than in 2010. It should not be surprising to learn that the risk of coverage errors was higher in 2020 than in 2010.

We examined 10 PSs that either directly or indirectly compared the error risks of various operations and activities for 2020 and 2010. Weighting these activities by their expected relative impacts on total error risks, we constructed an overall PS referred to as the SPS, which varies between 1 and 5, where 1 indicates a minimal increase in risk and 5 a maximal increase in risk. The results suggest that error risks, as measured by the SPS, increased relative to 2010 for 50 states and Washington, D.C. The SPS ranges from 1.21 (NE) to 4.47 (AK).

The SPS suggests that the risk of coverage error in 2020 is higher than in 2010. (What this says about the relative data quality is discussed further in our answer to Question 9.) However, while the analysis of PSs sheds light on potential error in the data collection for U.S. households, the strongest statement that can be made based on this analysis is the following: most of the critical activities considered in this report were exercised more frequently in 2020 than in 2010. To the extent an activity poses a coverage error risk, it follows that the risk of coverage errors also increased in 2020 for these activities. Despite these findings, there is no evidence that state-level counts used for apportionment purposes are of lower quality in 2020 than in 2010. Nor is there evidence that the apportionment count for any given state is in error.

A more conclusive assessment of 2020 census quality is probably best achieved by combining multiple approaches, such as PSs for critical activities and estimates of undercounts and overcounts from the Census Bureau's Post-Enumeration Survey and Demographic Analysis.

**2. Which critical activities, operations, or procedures were used in the 2020 census that were not used or were used much less frequently in 2010? Is there evidence from the PS analyses that these new census approaches increased or decreased error risk in 2020 compared to 2010?**

Over the past several decades, the Census Bureau has revised its approaches for improving census data collection and processing. Two examples are paid professional advertising for the 2000 census and the LUCA program aimed at improving the 2000 census Address List with local information. There is evidence that paid advertising improved initial census response and that LUCA added many addresses that were missing from the MAF. However, the LUCA program also greatly increased the workload for address de-duplication activity. This may have contributed to an overcount as estimated by the Census Bureau's 2000 coverage evaluation program.

A lesson from these two experiences is that innovations have both benefits and error risks. In 2010, problems with deploying new technology for field data collection and concerns about using the Internet for data collection delayed the adoption of these innovations. Although the 2010 census was judged a success in several areas, the use of the latest technology lagged. The decade that followed leading up to the 2020 census saw the Bureau make a big leap into new technologies and innovations in four areas to make the address list better, self-response easier, and the data collection more efficient:

a. Enhanced MAF updating

b. Self-response via the Internet

c. Promoting response without a census ID

d. Use of administrative records

*a. Enhanced MAF updating*

The Census Bureau began an effort to do more continuous updating of the address list through its Geographic Support System Initiative (GSS-I) in 2011 with a range of partners in tribal, state, and local governments, in addition to two yearly updates from the U.S. Postal Service. These efforts served to increase confidence in the MAF and led to the decision to replace the costly 100% pre-census address canvass with a combination of in-office and in-field address canvassing in preparation for the 2020 census. The Bureau "canvassed" about two-thirds of the nation using aerial photography, satellite imagery, and other local files to determine the validity of addresses for use in the census. These technological innovations were a new paradigm for the census, eclipsing the preparations for the 2010 census, which were largely confined to the LUCA program.

Of course, like the censuses of the past, innovation brought benefits and risks, and 2020 was no exception. Although the MAF was enhanced through the GSS-I, the selection of a subset of addresses for inclusion in the address list for the census was still based on a series of judgments about the quality of addresses, which were mostly untested, given the absence or cancellation of census tests because of funding shortfalls. Moreover, the potential error resulting from the 2020 in-office canvassing operation has not yet been assessed, although an assessment is forthcoming. Based on the available data, it is likely that these innovations increased error risk to some extent because of the large percentage of deleted and added addresses reflected in the MAF Revision PS. Moreover, the address "filtering" operation that created the initial census MAF also imparted error risks as it retained addresses that were unlikely to be valid, while excluding addresses that may have been valid.

*b. Self-response via the Internet*

Table 2 contrasts 2020 data collection to the 2010 data collection at a highly aggregated level. In 2020, a higher percentage of households self-responded, but a lower percentage responded by a household member in the NRFU phase. Adding these two percentages, data collection from the preferred methods (i.e., via a household member) was over 90% in 2020, compared to 93% in 2010. The Bureau used the Internet as the primary means of self-response for the first time in 2020 and more than 80% of self-respondents chose that option. Although self-response, particularly via the Internet, may be the most cost-effective and accurate data collection method, it produces error risks primarily because of multiple responses that then must be accurately resolved.

*Table 2. Percent Enumerated by Main Data Collection Methods: 2020 vs. 2010**

| Data Collection Method | 2020 Census | 2010 Census |
|---|---|---|
| Self-Response | 77.1 | 71.2 |
| Household Member Enumeration | 13.0 | 22.0 |
| Proxy Response | 5.4 | 6.4 |
| Administrative Records | 3.8 | 0.0 |
| Imputed Status/Counts | 0.7 | 0.4 |
| Total | 100.0 | 100.0 |

* Universe is all occupied HUs

*c. Promoting response with no ID*

In addition to response by Internet, the 2020 census promoted the submission of questionnaires without census IDs—a new approach that harnessed the ability of the MAF to accurately match non-ID returns to an address. About 22.2 million returns were submitted without a census ID, the majority of which were verified via automated match to the MAF. Of the remaining addresses about 9% could not be matched to the MAF, requiring more work on the part of the Census Bureau—in office and in the field—to determine their status (e.g., deletes, adds). Thus, a consequence of promoting non-ID responses is some level of risk associated with the large volume of nonmatching addresses that increased the workload for Bureau staff. Finally, an important item for future research is the impact of potential error in parsing addresses – many without standard features or sufficient detail – for determining whether two addresses are the same HU.

*d. Use of administrative records*

Although administrative records have been used in the past for selected populations, such as for certain types of GQs, this is the first time these were used to gauge occupancy status and to enumerate households. About 3.8% of all households were enumerated using administrative records in 2020. As discussed earlier, the use of administrative records has been controversial, especially as regards demographic characteristics. However, our sole focus here is the apportionment counts, and for counts, some data from 2010 Census evaluations do show that administrative records are superior to alternatives such as proxy responses and count imputations. As Table 2 shows, using proxy responses for counts in the NRFU declined between 2010 and 2020, from 6.4% to 5.4% of all households, while administrative records accounted for 3.8% in 2020, compared to zero in 2010.

It is likely that administrative records reduced reliance on proxy response and imputations in 2020. It is also likely that the observed increase in imputations—from 0.4% in 2010 to 0.7% in 2020—was minimized by the option to use administrative records. Still, the 0.3% increase in imputed status/counts is important given that it is the least preferred option for completing the census, with most of the increase the result of status imputation.

**3. The Census Bureau seeks to deploy its procedures in standardized fashion throughout the nation. However, conditions in each state can differ considerably regarding the base of addresses and difficulties surrounding the enumeration. How did the SPS vary across states? What does the range look like across states?**

Table 3 lists the 10 PSs, their means and minimum and maximum values for the 50 states and Washington, D.C. The last row of the table shows the same quantities for the SPS. The range is defined as the maximum PS minus the minimum PS. The relative range is the range divided by the absolute value of the mean PS. For almost all PSs, the relative range exceeds 1, which means that range is at least as large as the magnitude of the average of the 10 PSs across states. However, in two cases, it is less than 1: multiple responses (0.30) and administrative records use (0.95). For these two PSs there is not much variation across states. Risks for multiple responses and use of administrative records vary less as a result of these activities affecting all states in approximately the same way. In contrast, for other PSs (e.g., MAF Revisions), error risks vary substantially across the states.

*Table 3. Mean, Minimum, and Maximum of the 10 PSs*

| Process Statistic | Mean | Minimum | Maximum | Range | Relative Range |
|---|---|---|---|---|---|
| **1. MAF Revisions** | 11.3 | 5.9 | 26.6 | 20.7 | 1.8 |
| **2. Questionnaires Without ID not on MAF** | 9.0 | 4.3 | 21.6 | 17.3 | 1.9 |
| **3. Multiple Responses** | 17.2 | 15.0 | 20.2 | 5.2 | 0.3 |
| **4. Usual Residence at College** | 0.5 | -0.1 | 0.9 | 1.0 | 2.2 |
| **5. Responses Obtained by Proxy** | -0.5 | -2.1 | 1.7 | 3.9 | -8.4 |
| **6. Enumerations With Only a Population Count** | 0.5 | -1.3 | 1.6 | 2.8 | 6.0 |
| **7. Enumerations via Administrative Records** | 3.6 | 1.7 | 5.1 | 3.5 | 1.0 |
| **8. MAF Addresses Having Imputed Status** | 0.7 | 0.3 | 1.4 | 1.1 | 1.5 |
| **9. Occupied Housing Units With Imputed Population Counts** | -0.2 | -0.7 | 0.1 | 0.8 | -3.8 |
| **10. Group Quarters With Imputed Count** | 2.1 | 0.1 | 11.3 | 11.1 | 5.3 |
| **Summary Process Statistic** | 3.0 | 1.2 | 4.5 | 3.3 | 1.1 |

**4. What phases of the census contributed the most to SPS in states with the highest values of SPS?**

We can use the Phase Decomposition of SPS approach discussed in the Profiles section to address this question. As shown in Table 4, seven states have SPS values that exceed 4. In each case, the phase contributing most to SPS is self-response. By comparison, self-response contributes slightly more than 69% of SPS for the United States as a whole. MAF development and NRFU are the next largest contributors to total error risk. GQs contributes the least with less than 0.3% of SPS. For the United States as a whole, it is less than 0.13% of SPS.

*Table 4. Contribution of Census Phase to SPS for Top Seven PSs*

| State | Summary Process Statistic | MAF Contribution to SPS (%) | Self-Response Contribution to SPS (%) | NRFU Contribution to SPS (%) | Data Processing Contribution to SPS (%) | GQs Contribution to SPS (%) |
|---|---|---|---|---|---|---|
| AK | 4.5 | 40.5 | 53.6 | 4.1 | 1.5 | 0.2 |
| NJ | 4.4 | 7.4 | 67.6 | 23.3 | 1.6 | 0.0 |
| UT | 4.3 | 15.1 | 66.4 | 16.9 | 1.7 | 0.0 |
| NY | 4.3 | 20.8 | 61.8 | 14.7 | 2.7 | 0.0 |
| TX | 4.2 | 14.7 | 64.9 | 19.2 | 1.1 | 0.1 |
| MT | 4.1 | 40.1 | 58.3 | 1.1 | 0.3 | 0.1 |
| NM | 4.1 | 37.5 | 60.3 | 1.0 | 1.1 | 0.1 |

**5. What do the process statistics tell us about the accuracy of state numbers used for reapportionment?**

As emphasized throughout this report, neither the individual PSs nor the SPS alone can be used to make definitive statements about the accuracy of the apportionment counts. For example, although Alaska has the largest SPS among all states, the Post-Enumeration Survey could determine that net coverage error (i.e., overcounts minus undercounts) is quite small for that state. Likewise, Nebraska, having the smallest SPS at 1.21, may be shown to have larger net coverage error than a state with double that value. There are several reasons for this.

*a. Error risk is "opportunity" for error, not error*

Error risk reflects the number of opportunities for error. Thus, greater error risk simply means more opportunities for an error to occur. It is not known how many of these opportunities will result in an error (i.e., the error rate). For example, if we knew that for every 100 opportunities, one person is erroneously counted (i.e., a 1% error rate), then error risks could be converted to numbers of errors. However, because error rates are unknown, how data quality is affected by error risks is also unknown. Additionally, error rates are likely to vary a lot by PS; for example, the count imputation error rate is likely to be larger than the multiple responses resolution error rate. Since error rates are unknown, no statements regarding the magnitude of the errors in the apportionment counts by state or for the United States can be supported by the data in this report. However, because 2020 has greater error risk than 2010, it is more likely that the error in the state counts is larger (on average) in 2020 than it was in 2010. But, as the next point illustrates, the net coverage error may not be greater than in 2010.

*b. The accuracy of the apportionment counts depends on net coverage error and not on total number of errors.*

Some errors add to the apportionment counts (overcounts), while other errors subtract from them (undercounts). If the number of overcounted persons equals the number of undercounted persons for a state, then the state count would be perfectly accurate. For example, suppose a state was determined from our analysis to have 1 million opportunities for error. Furthermore, suppose that 1,000 of these opportunities resulted in undercounts and another 1,000 resulted in overcounts. Then, the apportionment count would be perfectly accurate because two errors would be offsetting. This is an important reason to avoid interpreting error risk as evidence of error in the apportionment counts. Knowing the error requires not only knowing the error rates for the 10 critical activities but also knowing their direction: positive or negative.

*c. The approach taken in this report is just one of many possible approaches, each potentially yielding somewhat different results and conclusions.*

The ASA Task Force Report provided specific guidance as to how the analysis should be approached and that guidance was followed in this report. Their report suggested about two dozen PSs or "quality indicators" as it referred to them for assessing 2020 census error risks. This report chose only 10 PSs—some suggested by the Task Force and others based on the consideration of activities that are most critical to census quality. Although these selected PSs are comprehensive and the analytical approach reflects the recommendations of the Task Force report, many other PSs could have been examined. Furthermore, the general approach adopted in this report (i.e., decomposing the census process into phases, using quintiles to compare risks across PSs, using weights to summarize these risks, using vetted methods to analyze the results) represents only one approach to the 2020 census quality assessment. Other approaches are possible and could produce different results and possibly different conclusions. Data for such experimentation are provided in Tables A1, A2, and A3 in the Appendix.

**6. There was a great deal of concern about the movement of college students related to the pandemic and its impact on the count. Do the PSs suggest any negative effects of these movements on the count?**

According to the Census Bureau's rules of where to count people, college students are supposed to be counted where they attend college. College students living in dormitories are included in the enumeration of people living in GQs, such as dormitories, halfway houses, and skilled nursing facilities. In addition, college students living off-campus were included in an early door-to-door operation carried out by mid-April to enumerate them before they left town at the end of the school year. However, because of the pandemic, most students left campus during March before that enumeration was completed. Questions remain on how successfully the Census Bureau managed to obtain school records from colleges and universities indicating where students were supposed to have been living on April 1. In addition, the risk that students may have been counted twice – at school and in the locations where they moved after campuses were shut down, such as at their parents' house – is a legitimate concern.

The question on the paper census form that asks, "Does this person usually live or stay somewhere else?" with an option for "Yes, for college" is shown in Figure 3. On the paper form, this question is asked of every person listed on the census questionnaire except the person completing the questionnaire. It is asked of everyone listed on the Internet or enumerator-assisted forms. (As previously described, the PS for the process of reassigning students enumerated in wrong locations is the difference between 2020 and 2010 for the ratio of the number of URCs to the number of HUs consisting of at least two persons expressed as a percentage.) Our results showed that the percentage increased in all states in 2020 except for Washington, D.C. Nationwide, about one percent of two-person HUs had at least one URC. The percentage of the population affected by this PS is small at the state level, but that may not be the case at substate levels.

*Figure 3. Census Question 2 That Assesses Usual Residence*

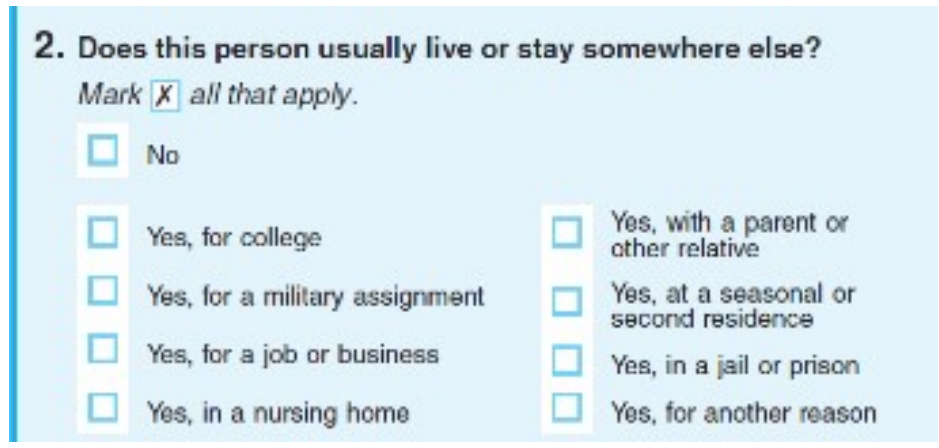
**2.** Does this person usually live or stay somewhere else?
*Mark [X] all that apply.*

☐ No

☐ Yes, for college
☐ Yes, for a military assignment
☐ Yes, for a job or business
☐ Yes, in a nursing home
☐ Yes, with a parent or other relative
☐ Yes, at a seasonal or second residence
☐ Yes, in a jail or prison
☐ Yes, for another reason

Similar to the GQs count imputation activity, URCs are highly localized and it is likely that the error in the URC reassignment activity is as well. Perhaps more importantly, the PS may reflect two related risks: (1) the reassignment of larger numbers of college students to alternate addresses in 2020 than in 2010, and (2) the risk of possible duplicate responses from college students who may be listed at two addresses. It may suggest the potential that college students may not have been properly identified as URCs on the census form and thus counted at the wrong address

**7. The GQs enumeration was greatly altered by the pandemic, given problems with access to facilities and changes in the approaches taken by the Bureau to collect data. What did the PSs tell us about the GQs enumeration?**

For the first time, the population counts for some GQs were imputed in 2020. As previously noted, this was likely the result of GQs enumeration difficulties caused by the pandemic. Although its impact at the state level may be small, the GQs count can be influential for substate statistics as evidenced by the wide variation in the percent of persons imputed by state. Locally, substantial coverage errors can result from the omission of a few large facilities and, although GQs count imputation may have reduced error risks, error risks remain because of the inherent inaccuracies of the count imputation process. Because GQ imputation is a new procedure that was required because of difficulties encountered in their enumeration, there are concerns regarding its implications for substate counts. An error risk evaluation for areas of high GQs concentrations is advisable; especially in Delaware and Mississippi where GQs count imputation rates were markedly higher than for other states. Local administrative records could be used as a resource to check the utility of the imputation methods (e.g., the kind that would have been deployed by the Federal State Cooperative on Population Estimates in the GQs count review that was unfortunately cancelled).

**8. The release of population estimates for states as of April 2020 allowed for comparison with the state census counts used for reapportionment. Is there any evidence from the PSs that states where population counts differed the most from their population estimates had higher levels of risk?**

The Population Estimates Program provides annual population estimates for geographic and demographic categories, including total population for states, cities, and towns. The most current population estimates that are available represent the resident population as of July 1, 2020, but the Census Bureau has issued estimates for April 1, 2020, for comparison with census counts by state. The population estimates have been carried forward from a 2010 census base, and likely have levels of uncertainty that make comparisons with the 2020 census counts challenging at best. Nonetheless, comparisons of the most recent population estimates available for states with the corresponding 2020 census counts may prove useful.

Of special interest are those states with large discrepancies, which may indicate inaccuracies for the 2020 census. What this analysis shows is that there is little if any relationship between the PSs and count-estimate discrepancies. The SPS has a correlation of just 0.17 with the estimates-count difference. Table 5 shows the top quintile of states on the two most dominant critical activities in this analysis: resolving multiple responses and MAF revisions. In both instances, the lists of top states on each PS present a disparate picture of differences between the population estimates and the census counts. There is no apparent pattern in magnitudes of the relative differences for these top-ranked states. This suggests that the relative difference may not be a reliable indicator of census error risks.

*Table 5. Relative Differences between Census and Population Estimates for Top States for Multiple Responses and MAF Revisions PSs*

| Top States for Multiple Responses | Relative Difference (Census—Population Estimate) |
|---|---|
| NM | 0.5 |
| AK | 0.2 |
| MT | 0.5 |
| NY | 4.2 |
| TX | -0.5 |
| NV | -0.8 |
| NH | 0.9 |
| CA | 0.3 |
| NJ | 4.5 |
| MN | 0.9 |
| UT | 1.0 |

| Top States for MAF Revisions | Relative Difference (Census—Population Estimate) |
|---|---|
| AK | 0.2 |
| WV | 0.3 |
| NM | 0.5 |
| MT | 0.5 |
| AR | -0.6 |
| WY | -0.9 |
| OK | -0.5 |
| ND | 1.8 |
| AL | 2.1 |
| HI | 3.2 |
| MS | -0.3 |

**9. What are the most important things we learned about the 2020 census from this analysis of the 10 PSs?**

a. Seven states—AK, MT, NJ, NM, NY, TX, UT—have SPSs exceeding 4, which suggests that these states have the highest risk for census error. These states have very different populations and range from mostly urban to mostly rural. Therefore, it is surprising they all rank in the top tier for SPS, which indicates that the error risks apply to very diverse populations and not only to either densely or sparsely populated areas of the country.

b. The Census Bureau made many important changes to ensure that the 2020 census counted everyone once and only once, and in the right place. These changes were designed to both reduce costs and increase the likelihood of response. However, as is usually the case with new procedures, these changes also increased error risks. Non-ID returns and multiple responses markedly increased in 2020 from 2010 levels. The address list required an unprecedented number of revisions, each carrying some, albeit small, error risk. Administrative records were employed that, in many ways, hugely benefited census data collection during the pandemic. However, although facilitating the counting process, administrative records increased uncertainty because their use can impart error. Unfortunately, the currently available data are insufficient for assessing the effect of these new approaches on coverage error.

c. Despite operating in one of the most challenging environments imaginable, the Census Bureau managed to successfully conduct the 2020 census. This is a considerable accomplishment and is testimony to the Bureau's laser focus on the execution of operations, many of which have stood the test of time and others that were new. Being a massive operational and engineering project, the Bureau must rely on data-driven decisions and well-tested methodologies. However, the lean fiscal environment that prevailed in intercensal years preceding 2020 curtailed some of the research and testing that would have been prudent considering the dramatic shift in census methodologies that occurred. As a result, the amount of error introduced by the critical activities identified in this report must be purely speculative. Perhaps in 2030, the Bureau will be better prepared to advance quality declarations that take error into account for any new, innovative methodologies it plans to roll out.

d. Many elements of the census are impressive in their design and execution. However, our assessment has especially brought a newfound appreciation of the complex process of building a census address list from the MAF. Without a complete and accurate MAF, other innovations and processes that are part of the execution of the census may become compromised, similar to a building with a poor foundation. Investments in research and evaluation on the potential errors related to decisions affecting the MAF in the 2020 census would go a long way as the Bureau looks forward to 2030.

e. At this time, little can be said about the causes of the differential between the population estimates published prior to the 2020 census and the 2020 census counts. The risk measures in this report do not explain why the census counts either fell short of or exceeded expectations. However, this result was anticipated because even the Post-Enumeration Survey results have not been able to explain gaps in actual versus expected counts in prior censuses. One reason is that the differential could be equally attributed to inaccuracies in the population estimates and to the census counts. It is important not to place much credence on differences between the census counts and the population estimates as an indicator of census error. While it may provide some information about error risk, it is not a reliable measure of census error.

## VII Acknowledgements

The authors acknowledge the support of the US Census Bureau, which provided the data and valuable information needed for this report. The authors also thank the American Statistical Association's 2020 Census Quality Indicators Task Force for their support and guidance.

## VIII References and Bibliography

# IX Appendix A

*Original Values for the 10 Process Statistic***s**

Table A1 provides the values of the 10 PSs prior to converting them to quintiles. The statistics in this table were cleared for public dissemination by the Census Bureau Disclosure Review Board. Clearance number CBDRB-FY21-DSSD007-0024.

Table A2 provides the values of the weights for the 10 PSs prior to scaling them to sum to 1. The statistics in this table were cleared for public dissemination by the Census Bureau Disclosure Review Board. Clearance number CBDRB-FY21-DSSD007-0024.

Table A3 provides the 2020 components for PSs that are based on the difference between 2020 and 2010. Subtracting the associated PS from the number in this table produces the corresponding PS component for 2010. The statistics in this table were cleared for public dissemination by the Census Bureau Disclosure Review Board. Clearance number CBDRB-FY21-DSSD007-0026.

*Table A1. Process Statistics by State*

| State | Relative Difference between the Census and the Population Estimates | 1. MAF Revisions (2020 Only) | 2. Questionnaires without ID not on MAF (2020 Only) | 3. Multiple Responses (2020 minus 2010) | 4. Usual Residence at College (2020 minus 2010) | 5. Responses Obtained by Proxy (2020 minus 2010) | 6. Enumerations with Only a Population Count (2020 minus 2010) | 7. Enumerations via Administrative Records (2020 Only) | 8. MAF Addresses Having Imputed Status (2020 minus 2010) | 9. Occupied Housing Units with Imputed Population Counts (2020 minus 2010) | 10. Group Quarters in TEA 1 & 6 with Imputed Count (2020 Only) |
|---|---|---|---|---|---|---|---|---|---|---|---|
| US | 0.62 | 9.22 | 8.79 | 17.53 | 0.53 | -0.35 | 0.53 | 3.84 | 0.74 | -0.22 | 2.01 |
| AL | 2.11 | 14.38 | 7.21 | 15.18 | 0.60 | -1.72 | -0.26 | 3.51 | 0.90 | -0.44 | 4.98 |
| AK | 0.18 | 26.55 | 21.55 | 20.14 | 0.18 | -0.62 | 1.30 | 2.11 | 0.81 | -0.25 | 4.76 |
| AZ | -3.28 | 10.65 | 11.25 | 18.13 | 0.35 | -0.79 | -1.02 | 3.41 | 0.48 | -0.33 | 3.19 |
| AR | -0.60 | 16.01 | 7.52 | 16.69 | 0.20 | -1.37 | 0.54 | 3.36 | 0.71 | -0.17 | 3.57 |
| CA | 0.30 | 8.82 | 6.47 | 18.82 | 0.78 | -0.11 | 0.79 | 3.75 | 0.72 | -0.07 | 2.00 |
| CO | -0.42 | 9.95 | 8.16 | 18.09 | 0.54 | -1.11 | -0.84 | 3.55 | 0.66 | -0.44 | 1.10 |
| CT | 1.25 | 7.56 | 11.58 | 17.41 | 0.66 | 0.21 | 1.06 | 4.27 | 0.84 | 0.01 | 0.50 |
| DE | 0.51 | 6.24 | 5.38 | 16.68 | 0.62 | -1.78 | -0.08 | 4.87 | 0.25 | 0.02 | 11.26 |
| DC | -3.18 | 11.51 | 7.16 | 16.52 | -0.11 | -0.98 | 1.26 | 4.98 | 0.74 | -0.67 | 0.26 |
| FL | -0.69 | 7.91 | 7.90 | 16.97 | 0.42 | -0.28 | 0.15 | 4.43 | 0.64 | -0.38 | 3.53 |
| GA | 0.15 | 12.18 | 5.88 | 17.88 | 0.68 | -0.87 | -0.61 | 4.32 | 0.71 | -0.60 | 1.99 |
| HI | 3.17 | 14.28 | 14.09 | 15.66 | 0.67 | -0.01 | -1.25 | 1.65 | 1.00 | -0.14 | 0.63 |
| ID | 1.15 | 11.17 | 8.53 | 15.95 | 0.13 | -0.64 | -0.38 | 2.51 | 0.53 | -0.27 | 0.50 |
| IL | 1.56 | 8.71 | 12.03 | 17.04 | 0.64 | 0.09 | 1.10 | 3.64 | 0.73 | -0.15 | 1.04 |
| IN | 0.49 | 7.25 | 5.51 | 15.43 | 0.37 | -0.21 | 0.91 | 3.62 | 0.60 | -0.01 | 0.64 |
| IA | 0.83 | 7.27 | 5.14 | 17.21 | 0.76 | 0.29 | 1.09 | 4.23 | 0.54 | -0.01 | 2.25 |
| KS | 0.78 | 7.98 | 4.95 | 16.04 | 0.59 | 0.69 | 1.20 | 3.05 | 0.66 | 0.01 | 0.57 |
| KY | 0.62 | 11.91 | 7.93 | 15.45 | 0.23 | -1.74 | 0.13 | 3.18 | 0.69 | -0.08 | 2.17 |
| LA | 0.15 | 12.77 | 6.45 | 17.20 | 0.28 | -1.90 | -0.19 | 5.12 | 1.36 | -0.49 | 4.34 |
| ME | 0.94 | 11.90 | 7.79 | 14.98 | 0.32 | 0.03 | 0.85 | 4.12 | 0.80 | -0.05 | 5.16 |
| MD | 1.94 | 5.87 | 4.28 | 17.78 | 0.55 | -0.31 | 0.79 | 4.04 | 0.55 | -0.18 | 1.23 |
| MA | 1.91 | 10.18 | 10.60 | 17.46 | 0.45 | 0.00 | 1.18 | 4.19 | 1.06 | 0.05 | 0.58 |
| MI | 1.01 | 7.89 | 5.50 | 17.39 | 0.76 | -0.87 | 0.51 | 3.36 | 0.69 | -0.12 | 4.81 |
| MN | 0.90 | 6.48 | 6.08 | 18.34 | 0.79 | -0.03 | 0.95 | 2.85 | 0.60 | -0.11 | 0.32 |
| MS | -0.34 | 14.23 | 7.25 | 15.25 | 0.20 | -1.89 | 0.87 | 3.89 | 0.84 | -0.47 | 6.81 |
| MO | 0.05 | 13.19 | 6.46 | 16.24 | 0.64 | -0.91 | 0.62 | 3.61 | 0.75 | -0.12 | 3.95 |
| MT | 0.54 | 19.30 | 12.30 | 19.88 | 0.37 | -0.42 | -0.60 | 2.73 | 0.56 | -0.10 | 2.01 |
| NE | 1.25 | 7.79 | 5.85 | 15.73 | 0.52 | -0.07 | 0.68 | 3.47 | 0.62 | -0.04 | 0.41 |
| NV | -0.76 | 7.79 | 6.63 | 18.97 | 0.25 | -1.73 | -1.22 | 4.39 | 0.55 | -0.41 | 1.72 |
| NH | 0.88 | 8.79 | 7.45 | 18.87 | 0.55 | -0.35 | 0.88 | 4.46 | 0.69 | 0.00 | 0.13 |
| NJ | 4.48 | 8.33 | 12.71 | 18.52 | 0.58 | 0.14 | 1.37 | 4.09 | 0.78 | -0.28 | 0.68 |
| NM | 0.54 | 19.68 | 16.42 | 20.15 | 0.39 | -1.70 | -0.62 | 2.68 | 0.71 | -0.71 | 3.15 |
| NY | 4.23 | 13.32 | 15.76 | 19.61 | 0.21 | -0.29 | 1.59 | 4.37 | 1.19 | -0.23 | 0.79 |
| NC | -1.35 | 10.55 | 5.29 | 17.35 | 0.58 | -1.17 | -0.11 | 4.52 | 0.67 | -0.43 | 3.07 |
| ND | 1.81 | 14.98 | 11.02 | 16.76 | 0.39 | 0.28 | 1.15 | 2.90 | 0.77 | -0.07 | 0.32 |
| OH | 0.85 | 7.19 | 6.10 | 16.97 | 0.56 | -0.07 | 0.70 | 4.07 | 0.65 | -0.06 | 1.14 |
| OK | -0.46 | 15.00 | 9.71 | 16.91 | 0.23 | -0.42 | 0.63 | 2.85 | 0.65 | -0.05 | 0.53 |
| OR | 0.00 | 10.20 | 7.53 | 17.23 | 0.44 | -0.11 | 0.80 | 2.97 | 0.63 | -0.24 | 1.64 |
| PA | 1.63 | 8.83 | 8.76 | 15.90 | 0.62 | 0.34 | 0.93 | 4.04 | 0.73 | -0.08 | 1.60 |
| RI | 3.72 | 7.83 | 15.49 | 17.10 | 0.17 | 1.74 | 0.80 | 5.09 | 1.03 | 0.01 | 0.21 |
| SC | -1.68 | 12.56 | 4.90 | 15.38 | 0.25 | -0.28 | 0.30 | 4.26 | 0.69 | -0.33 | 3.49 |
| SD | -0.56 | 12.89 | 19.20 | 15.94 | 0.51 | -0.37 | 0.29 | 3.10 | 0.63 | -0.34 | 1.00 |
| TN | 0.51 | 10.29 | 5.15 | 15.70 | 0.43 | -1.07 | 0.35 | 3.94 | 0.67 | -0.32 | 3.62 |
| TX | -0.48 | 11.25 | 6.97 | 19.04 | 0.59 | 0.07 | 0.06 | 3.97 | 0.70 | -0.39 | 1.90 |
| UT | 0.99 | 11.01 | 13.59 | 18.32 | 0.17 | 0.77 | 1.13 | 2.54 | 0.80 | -0.03 | 0.55 |
| VT | 3.09 | 13.63 | 9.24 | 16.67 | 0.40 | 0.69 | 1.41 | 3.09 | 0.85 | 0.05 | 0.70 |
| VA | 0.51 | 6.75 | 4.80 | 17.65 | 0.89 | -0.49 | 0.13 | 3.64 | 0.60 | -0.13 | 0.71 |
| WA | 0.35 | 7.80 | 8.37 | 16.60 | 0.47 | -0.38 | 0.88 | 3.25 | 0.81 | -0.20 | 2.72 |
| WV | 0.31 | 25.58 | 15.07 | 16.52 | 0.13 | -2.11 | 0.62 | 2.74 | 0.75 | 0.02 | 0.53 |
| WI | 1.03 | 7.07 | 6.81 | 16.31 | 0.77 | 0.21 | 1.01 | 3.17 | 0.67 | -0.09 | 1.23 |
| WY | -0.89 | 15.36 | 10.68 | 16.55 | 0.31 | 0.24 | 0.04 | 3.38 | 0.67 | -0.31 | 1.92 |

*Table A2. Process Statistic Weights by State*

| State | 1. MAF Revisions (2020 Only) | 2. Questionnaires without ID not on MAF (2020 Only) | 3. Multiple Responses (2020 minus 2010) | 4. Usual Residence at College (2020 minus 2010) | 5. Responses Obtained by Proxy (2020 minus 2010) | 6. Enumerations with Only a Population Count (2020 minus 2010) | 7. Enumerations via Administrative Records (2020 Only) | 8. MAF Addresses Having Imputed Status (2020 minus 2010) | 9. Occupied Housing Units with Imputed Population Counts (2020 minus 2010) | 10. Group Quarters in TEA 1 & 6 with Imputed Count (2020 Only) |
|---|---|---|---|---|---|---|---|---|---|---|
| AL | 14.38 | 0.76 | 25.44 | 1.05 | 4.34 | 2.01 | 3.51 | 1.09 | 0.10 | 0.13 |
| AK | 26.55 | 2.50 | 32.55 | 0.60 | 5.52 | 2.29 | 2.11 | 1.01 | 0.05 | 0.16 |
| AZ | 10.65 | 0.93 | 27.39 | 0.69 | 5.87 | 2.10 | 3.41 | 0.68 | 0.03 | 0.07 |
| AR | 16.01 | 0.69 | 24.81 | 0.62 | 3.99 | 1.89 | 3.36 | 0.91 | 0.05 | 0.10 |
| CA | 8.82 | 0.54 | 28.56 | 1.21 | 4.42 | 2.44 | 3.75 | 0.81 | 0.05 | 0.05 |
| CO | 9.95 | 0.59 | 25.83 | 0.88 | 4.71 | 2.18 | 3.55 | 0.77 | 0.06 | 0.02 |
| CT | 7.56 | 0.79 | 26.15 | 1.30 | 4.48 | 2.35 | 4.27 | 0.98 | 0.05 | 0.01 |
| DE | 6.24 | 0.41 | 25.65 | 1.10 | 4.29 | 2.38 | 4.87 | 0.64 | 0.08 | 0.26 |
| DC | 11.51 | 0.76 | 29.06 | 0.77 | 6.34 | 3.31 | 4.98 | 1.01 | 0.10 | 0.02 |
| FL | 7.91 | 0.65 | 26.39 | 0.84 | 5.17 | 2.46 | 4.43 | 0.81 | 0.06 | 0.08 |
| GA | 12.18 | 0.54 | 28.15 | 1.26 | 5.28 | 2.11 | 4.32 | 0.90 | 0.12 | 0.05 |
| HI | 14.28 | 1.13 | 29.96 | 1.01 | 5.66 | 1.69 | 1.65 | 1.16 | 0.04 | 0.02 |
| ID | 11.17 | 0.53 | 24.60 | 0.73 | 3.79 | 1.89 | 2.51 | 0.71 | 0.06 | 0.01 |
| IL | 8.71 | 0.98 | 25.50 | 1.17 | 4.56 | 2.69 | 3.64 | 0.85 | 0.05 | 0.02 |
| IN | 7.25 | 0.37 | 22.92 | 0.82 | 4.04 | 2.04 | 3.62 | 0.70 | 0.04 | 0.02 |
| IA | 7.27 | 0.28 | 23.43 | 1.05 | 3.95 | 1.87 | 4.23 | 0.61 | 0.05 | 0.07 |
| KS | 7.98 | 0.31 | 23.56 | 0.94 | 4.30 | 1.85 | 3.05 | 0.74 | 0.06 | 0.02 |
| KY | 11.91 | 0.77 | 23.58 | 0.65 | 3.94 | 1.87 | 3.18 | 0.90 | 0.07 | 0.06 |
| LA | 12.77 | 0.69 | 26.84 | 0.87 | 4.56 | 2.27 | 5.12 | 1.53 | 0.17 | 0.12 |
| ME | 11.90 | 0.62 | 22.87 | 0.79 | 4.01 | 1.85 | 4.12 | 0.93 | 0.04 | 0.14 |
| MD | 5.87 | 0.34 | 27.31 | 1.24 | 4.30 | 2.44 | 4.04 | 0.71 | 0.06 | 0.03 |
| MA | 10.18 | 0.80 | 26.71 | 1.26 | 4.40 | 2.27 | 4.19 | 1.17 | 0.11 | 0.02 |
| MI | 7.89 | 0.45 | 24.98 | 1.16 | 3.43 | 2.08 | 3.36 | 0.81 | 0.05 | 0.11 |
| MN | 6.48 | 0.35 | 25.17 | 1.17 | 3.40 | 1.83 | 2.85 | 0.68 | 0.03 | 0.01 |
| MS | 14.23 | 0.73 | 25.70 | 0.99 | 3.46 | 2.09 | 3.89 | 1.04 | 0.12 | 0.22 |
| MO | 13.19 | 0.41 | 23.76 | 0.94 | 3.74 | 2.26 | 3.61 | 0.86 | 0.07 | 0.11 |
| MT | 19.30 | 1.19 | 26.54 | 0.72 | 4.62 | 2.03 | 2.73 | 0.82 | 0.08 | 0.05 |
| NE | 7.79 | 0.33 | 22.49 | 0.91 | 3.89 | 1.95 | 3.47 | 0.71 | 0.02 | 0.01 |
| NV | 7.79 | 0.59 | 27.57 | 0.53 | 5.69 | 2.65 | 4.39 | 0.78 | 0.09 | 0.02 |
| NH | 8.79 | 0.55 | 26.56 | 1.17 | 3.70 | 1.91 | 4.46 | 0.80 | 0.05 | 0.00 |
| NJ | 8.33 | 1.06 | 28.18 | 1.40 | 4.59 | 2.60 | 4.09 | 0.92 | 0.06 | 0.01 |
| NM | 19.68 | 2.03 | 29.31 | 0.72 | 5.47 | 2.51 | 2.68 | 0.98 | 0.06 | 0.06 |
| NY | 13.32 | 1.64 | 29.86 | 0.98 | 5.44 | 3.17 | 4.37 | 1.39 | 0.09 | 0.02 |
| NC | 10.55 | 0.46 | 26.85 | 1.02 | 4.71 | 2.19 | 4.52 | 0.89 | 0.09 | 0.08 |
| ND | 14.98 | 1.05 | 23.13 | 0.78 | 4.40 | 2.03 | 2.90 | 0.89 | 0.03 | 0.01 |
| OH | 7.19 | 0.43 | 24.32 | 0.95 | 4.33 | 2.08 | 4.07 | 0.73 | 0.05 | 0.03 |
| OK | 15.00 | 0.88 | 25.66 | 0.70 | 5.23 | 1.87 | 2.85 | 0.82 | 0.03 | 0.02 |
| OR | 10.20 | 0.50 | 25.33 | 0.77 | 4.36 | 2.00 | 2.97 | 0.74 | 0.06 | 0.04 |
| PA | 8.83 | 0.58 | 24.18 | 1.20 | 4.44 | 2.03 | 4.04 | 0.86 | 0.07 | 0.05 |
| RI | 7.83 | 1.07 | 25.57 | 0.87 | 6.65 | 2.46 | 5.09 | 1.15 | 0.10 | 0.01 |
| SC | 12.56 | 0.40 | 25.59 | 0.85 | 4.79 | 1.94 | 4.26 | 0.91 | 0.08 | 0.09 |
| SD | 12.89 | 1.23 | 22.92 | 0.94 | 4.17 | 1.77 | 3.10 | 0.79 | 0.03 | 0.04 |
| TN | 10.29 | 0.40 | 24.26 | 0.85 | 4.44 | 2.12 | 3.94 | 0.80 | 0.06 | 0.08 |
| TX | 11.25 | 0.67 | 28.64 | 1.07 | 5.84 | 2.34 | 3.97 | 0.85 | 0.05 | 0.04 |
| UT | 11.01 | 0.94 | 27.90 | 0.87 | 4.35 | 2.52 | 2.54 | 0.90 | 0.05 | 0.01 |
| VT | 13.63 | 0.88 | 24.87 | 0.96 | 4.08 | 1.79 | 3.09 | 0.99 | 0.06 | 0.03 |
| VA | 6.75 | 0.37 | 26.58 | 1.30 | 4.17 | 2.01 | 3.64 | 0.72 | 0.04 | 0.02 |
| WA | 7.80 | 0.61 | 25.41 | 0.82 | 4.04 | 2.23 | 3.25 | 0.92 | 0.04 | 0.06 |
| WV | 25.58 | 1.78 | 24.96 | 0.65 | 4.44 | 1.69 | 2.74 | 0.99 | 0.07 | 0.02 |
| WI | 7.07 | 0.36 | 22.72 | 1.04 | 3.64 | 2.04 | 3.17 | 0.78 | 0.05 | 0.03 |
| WY | 15.36 | 1.02 | 24.83 | 0.75 | 5.16 | 2.25 | 3.38 | 0.89 | 0.02 | 0.04 |

*Table A3. 2020 Components for 2020 - 2010 Process Statistics*

| State | 3. Percent of Occupied HUs Having Multiple Responses | 4. Percent of Occupied HUs with Usual Residence at College | 5. Percent of Persons whose Responses were Obtained by Proxy | 6. Percent of Occupied HUs with Only a Population Count | 8. Percent of MAF Addresses with Imputed Status | 9. Percent of Occupied HUs with Imputed Population Counts |
|---|---|---|---|---|---|---|
| US | 26.47 | 1.45 | 4.65 | 2.28 | 0.88 | 0.06 |
| AL | 25.44 | 1.51 | 4.34 | 2.01 | 1.09 | 0.10 |
| AK | 32.55 | 0.84 | 5.52 | 2.29 | 1.01 | 0.05 |
| AZ | 27.39 | 0.95 | 5.87 | 2.10 | 0.68 | 0.03 |
| AR | 24.81 | 0.90 | 3.99 | 1.89 | 0.91 | 0.05 |
| CA | 28.56 | 1.62 | 4.42 | 2.44 | 0.81 | 0.05 |
| CO | 25.83 | 1.24 | 4.71 | 2.18 | 0.77 | 0.06 |
| CT | 26.15 | 1.87 | 4.48 | 2.35 | 0.98 | 0.05 |
| DE | 25.65 | 1.54 | 4.29 | 2.38 | 0.64 | 0.08 |
| DC | 29.06 | 1.40 | 6.34 | 3.31 | 1.01 | 0.10 |
| FL | 26.39 | 1.19 | 5.17 | 2.46 | 0.81 | 0.07 |
| GA | 28.15 | 1.76 | 5.28 | 2.11 | 0.90 | 0.12 |
| HI | 29.96 | 1.36 | 5.66 | 1.69 | 1.16 | 0.04 |
| ID | 24.60 | 0.98 | 3.79 | 1.89 | 0.71 | 0.06 |
| IL | 25.50 | 1.71 | 4.56 | 2.69 | 0.85 | 0.05 |
| IN | 22.92 | 1.17 | 4.04 | 2.04 | 0.70 | 0.04 |
| IA | 23.43 | 1.53 | 3.95 | 1.87 | 0.61 | 0.05 |
| KS | 23.56 | 1.34 | 4.30 | 1.85 | 0.74 | 0.06 |
| KY | 23.58 | 0.93 | 3.94 | 1.87 | 0.90 | 0.08 |
| LA | 26.84 | 1.26 | 4.56 | 2.27 | 1.53 | 0.17 |
| ME | 22.87 | 1.15 | 4.01 | 1.85 | 0.93 | 0.04 |
| MD | 27.31 | 1.74 | 4.30 | 2.44 | 0.71 | 0.07 |
| MA | 26.71 | 1.81 | 4.40 | 2.27 | 1.17 | 0.11 |
| MI | 24.98 | 1.68 | 3.43 | 2.08 | 0.81 | 0.05 |
| MN | 25.17 | 1.67 | 3.40 | 1.83 | 0.68 | 0.03 |
| MS | 25.70 | 1.44 | 3.46 | 2.09 | 1.04 | 0.12 |
| MO | 23.76 | 1.38 | 3.74 | 2.26 | 0.86 | 0.07 |
| MT | 26.54 | 1.06 | 4.62 | 2.03 | 0.82 | 0.08 |
| NE | 22.49 | 1.32 | 3.89 | 1.95 | 0.71 | 0.02 |
| NV | 27.57 | 0.74 | 5.69 | 2.65 | 0.78 | 0.09 |
| NH | 26.56 | 1.64 | 3.70 | 1.91 | 0.80 | 0.05 |
| NJ | 28.18 | 1.93 | 4.59 | 2.60 | 0.92 | 0.06 |
| NM | 29.31 | 1.05 | 5.47 | 2.51 | 0.99 | 0.06 |
| NY | 29.86 | 1.46 | 5.44 | 3.17 | 1.39 | 0.09 |
| NC | 26.85 | 1.46 | 4.71 | 2.19 | 0.89 | 0.09 |
| ND | 23.13 | 1.18 | 4.40 | 2.03 | 0.89 | 0.03 |
| OH | 24.32 | 1.40 | 4.33 | 2.08 | 0.73 | 0.05 |
| OK | 25.66 | 1.00 | 5.23 | 1.87 | 0.82 | 0.03 |
| OR | 25.33 | 1.08 | 4.36 | 2.00 | 0.74 | 0.06 |
| PA | 24.18 | 1.73 | 4.44 | 2.03 | 0.86 | 0.07 |
| RI | 25.57 | 1.29 | 6.65 | 2.46 | 1.15 | 0.10 |
| SC | 25.59 | 1.20 | 4.79 | 1.94 | 0.91 | 0.08 |
| SD | 22.92 | 1.37 | 4.17 | 1.77 | 0.79 | 0.03 |
| TN | 24.26 | 1.21 | 4.44 | 2.12 | 0.80 | 0.06 |
| TX | 28.64 | 1.46 | 5.84 | 2.34 | 0.85 | 0.05 |
| UT | 27.90 | 1.12 | 4.35 | 2.52 | 0.90 | 0.05 |
| VT | 24.87 | 1.41 | 4.08 | 1.79 | 0.99 | 0.06 |
| VA | 26.58 | 1.82 | 4.17 | 2.01 | 0.72 | 0.04 |
| WA | 25.41 | 1.15 | 4.04 | 2.23 | 0.93 | 0.04 |
| WV | 24.96 | 0.95 | 4.44 | 1.69 | 0.99 | 0.07 |
| WI | 22.72 | 1.51 | 3.64 | 2.04 | 0.78 | 0.05 |
| WY | 24.83 | 1.09 | 5.16 | 2.25 | 0.89 | 0.02 |